%% file: paper.tex
\documentclass{iopart}
\usepackage{graphicx}

\begin{document}
\title{Casimir torque}
\author{Jos\'e C. Torres-Guzm\'an$^{1,2}$ and W. Luis Moch\'an$^1$}
\address{$^1$Centro de Ciencias F\'{\i}sicas, Universidad Nacional
Aut\'onoma de M\'exico, Apartado Postal 48-3, 62251 Cuernavaca,
Morelos, M\'exico}
\address{$^2$Facultad de Ciencias, Universidad Aut\'onoma del Estado
de Morelos, Avenida Universidad 1001, 62221 Cuernavaca, Morelos, M\'exico}
\ead{\mailto{JCTG <torres@fis.unam.mx>}, \mailto{WLM <mochan@fis.unam.mx>}}
\begin{abstract}
We develop a formalism for the calculation of the flow of angular
momentum carried by the fluctuating electromagnetic field within a
cavity bounded by two flat anisotropic materials. By generalizing a procedure
employed recently for the calculation of the Casimir force between
arbitrary materials, we obtain an expression for the torque between
anisotropic plates in terms of their reflection amplitude
matrices. We evaluate the torque in 1D for ideal and realistic model
materials.
\end{abstract}
\submitto{\JPA}
\section{Introduction}

In the last decade the Casimir effect \cite{casimir} has received considerable
attention,  as
the recently attained high experimental accuracy has permitted
detailed tests of  
theoretical predictions \cite{1,2,3,4,5,6,7,8}.  This, in
turn, has stimulated a growing interest in fundamental aspects
of the 
vacuum field.  The study of vacuum forces between realistic materials
was pioneered by Lifshitz \cite{9}, who considered local
homogeneous materials whose fluctuating currents were the sources of
the fluctuating electromagnetic field and whose correlations were
related to the dielectric response of the materials. One of the
limitations of 
the Lifshitz theory is the requirement of a
definite microscopic model of matter which has to be solved
simultaneously with the electromagnetic field equations. Thus, the
applicability of the results seem to be limited by the generality of
that initial model. In particular, Lifshitz results were developed for
isotropic materials. Nevertheless, the vacuum energy has been
calculated for cavities bounded by anisotropic materials, first in the
non-retarded limit \cite{10} and later for arbitrary
distances \cite{11}, resulting in a torque whenever 
the optical axes of the plates are not aligned with each other. An alternative
derivation of the Casimir torque in the 1D case has been developed
\cite{12} starting from the angular momentum flux carried by the 
field. Analytical formulae have been found in the retarded limit when the
anisotropy is small \cite{13}. Numerical calculations have also been
performed for materials with 
a small anisotropy and it has been shown that the torque may be large
enough to be experimentally measurable in several novel experimental
configurations \cite{14}. 

In the previous theoretical works essential assumptions
about the dielectric properties of the plates were done from the onset
in order to derive expressions for the Casimir torque. However, recent
works \cite{lambrecht,15} have shown that if the
theory is set up in terms of the reflection coefficients of
the media, it is possible to decouple the calculation of the Casimir
force from the calculation of the dielectric response of the
materials. The so called scattering approach has permitted the calculation
of the force for a wide class of systems simply by plugging into the
resulting formulae the appropriate reflection amplitudes or surface
impedances. Thus, 
transparent and opaque, local and non-local, infinite and finite, 
homogeneous and heterogeneous systems may be treated in the same
footing. In the present paper we generalize the scattering approach to
account for anisotropy as well. 
We present a new derivation of the Casimir
torque between plates with arbitrary dielectric
properties characterized by their anisotropic optical
coefficients. For simplicity, in this paper we focus our attention on one
dimensional systems, although our approach is also applicable to 3D
\cite{unpub}. We present results for both ideal systems, 
for which analytical formulae are obtained, and realistic dichroic
systems.

\section{Scattering approach}

\begin{figure}
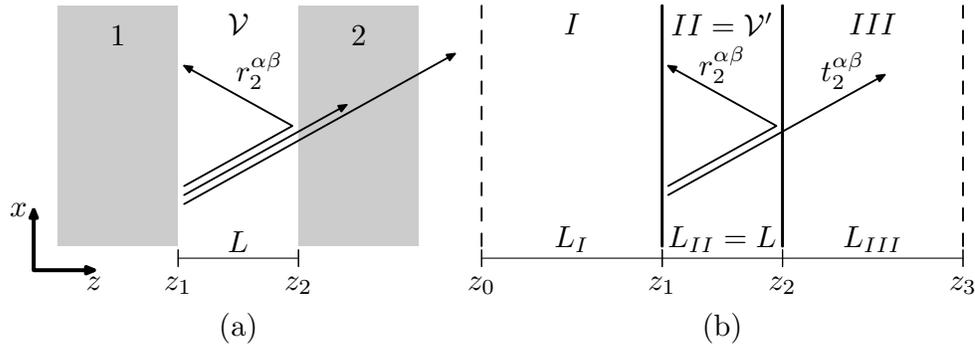

\centering{\includegraphics{paper.2} \includegraphics{paper.3}}
\caption{\label{scattering}(a) Vacuum cavity  $\mathcal V$ of
width $L$ 
bounded by two arbitrary material slabs (1 and 2) with surfaces at
$z_1$ and $z_2$. A photon with polarization $\hat e_i^\beta$ propagating
within the cavity is either reflected coherently with
amplitude and polarization $r_2 \hat e_r^\alpha=r_2^{\alpha\beta}
e^\beta$, or lost from the cavity with probability $1-|r|^2$. 
(b)Fictitious system made up three empty
regions $I$, $II$, and $III$, bounded by perfect mirrors at
$z_0$ and $z_3$ and with infinitely thin
sheets at $z_1$ and $z_2$ whose
reflection amplitudes 
$r_a^{\alpha\beta}$ are identical to those of the real system
and whose transmission amplitudes
$t_a^{\alpha\beta}$ are such that energy is conserved with no
absorption whatsoever. The field within the fictitious cavity $\mathcal V'$
(region $II$) is the same as within the real cavity $\mathcal V$.}  
\end{figure}

To calculate the torque we follow the scattering approach
\cite{lambrecht,15}, illustrated by Fig.~\ref{scattering}(a).
A photon within the cavity may be described by its amplitude $\mathcal
E$,
frequency $\omega$, wavevector $\vec k=(\vec Q, \pm q)$ and polarization
$\hat e^\alpha$.
We may
chose the independent polarizations as $\alpha=s,p$ (or equivalently, TE and TM). 
When a photon with
polarization $\hat e^\alpha_i$ is incident upon the surface, say, of
medium 2 at $z_2$, it is either reflected coherently with an amplitude and
polarization proportional to
$r_2 \hat e_r^\alpha\equiv r_2^{\alpha\beta} \hat e_i^\beta$ (sum
implied), or 
transmitted with a 
probability $1-|r_2|^2$. Here, $r_a^{\alpha\beta}$ is a $2\times 2$
matrix that describes the reflection amplitude of medium $a=1,2$. For
isotropic media, $r_a^{\alpha\beta}$ may be taken as a scalar whenever
the incoming field has $s$ or $p$ polarization, but that
separation is not possible in general when the media are
anisotropic. We remark that $r_a^{\alpha\beta}$ is defined to be the
complete reflection amplitude of medium $a$, not only that of its
front surface. For instance, if medium $a$ were 
a thin film or a layered system, the multiple reflections within $a$
are to be incorporated into $r_a^{\alpha\beta}$. Thus, if the photon is
not coherently reflected, it must necessarily be absorbed within $a$
or else be transmitted into the empty space  beyond. The principle of
detailed balance implies that in
thermodynamic equilibrium, for every photon that is not reflected and
is therefore lost from the
cavity either through absorption or transmission, an equivalent photon
is incoherently launched into the cavity, either being radiated by the
absorbing medium, or else, arriving from the vacuum beyond and
being transmitted into the cavity. In any case, the probability that a
photon with wavevector $(\vec Q,-q)$ and polarization $\hat
e_r^\alpha$ arrives into the cavity from medium 2 with no phase
relation to the lost photon is proportional to
$1-|r_2|^2$. Similar statements apply to medium 1. 

From the previous discussion, it follows that in equilibrium 
the properties of the radiation field within the cavity $\mathcal
V$ depend on the cavity walls only through their optical coefficients
$r_a^{\alpha\beta}$. Equivalently, the cavity
radiation is completely determined by the exact surface impedance
$Z_a^{\mu\nu}$ defined through $[\hat n_a\times (\hat n_a \times \vec
E_a)]^\mu=Z_a^{\mu\nu} (\hat n_a\times \vec H_a)^\nu$ (sum implied), where
$\hat n_a$ denotes the outgoing unit normal of surface $a$, $\vec E_a$
and $\vec H_a$ are the total electric and magnetic fields at $z_a$ and
$\mu,\nu=x,y$ denote  Cartesian coordinates along the walls. Thus, the
electromagnetic radiation within the real cavity $\mathcal V$ would be
identical to that within a fictitious cavity $\mathcal V'$ bounded by
infinitely thin sheets at $z_1$ and $z_2$, provided their reflection
amplitudes $r_a^{\alpha\beta}$ are chosen to be equal to those of the
walls of $\mathcal V$. Their transmission amplitudes
$t_a^{\alpha\beta}$ may then be chosen in 
order to guarantee energy conservation with no absorption whatsoever
of electromagnetic energy (Fig. \ref{scattering}(b)).
As there is no absorption in the fictitious system, there is no
excitation of material degrees of freedom and the normal modes of the
electromagnetic field form a complete orthogonal basis of the
corresponding Hilbert space, allowing the use of well developed
quantum-mechanical procedures for the calculation of the field
properties. Contrariwise, in the real system the electromagnetic
energy is absorbed, probably exciting electronic or vibrational
transitions, so that the 
problem cannot be treated quantum mechanically without incorporating
the electronic and/or vibrational degrees of freedom into the
calculation, which would require 
in turn the use of a microscopic model of the material. 

In Fig. \ref{scattering}(b) we have included two perfect mirrors at
positions $z_0$ and $z_3$ in order to quantize and count the normal
modes of the system. The photons that are reflected at $z_0$ and $z_3$
and are transmitted back into $\mathcal V'$ mimic the
photons injected into the real cavity $\mathcal V$ in order to restore
thermal equilibrium, and in the limit $L_I, L_{III}\to \infty$ their
phase is so large and so rapidly varying with $\omega$ that it
effectively bears no relation with the phase of the photons lost from
the cavity.

Using the
scattering approach we can treat dissipationless, homogeneous,
isotropic, local, sharp media on the same footing as dissipative,
inhomogeneous (layered systems, superlattices, photonic structures),
chiral, spatially dispersive materials with a smooth selvedge. In
particular, we can treat anisotropic systems.

\section{Torque in 1D}

We consider a finite beam propagating within $\mathcal V'$ along $\pm z$,
\begin{equation}\label{E}
\vec E(\vec r, t)=E_r \hat e_r e^{i(qz-\omega t)}+E_l \hat e_l
e^{-i(qz+\omega t)},
\end{equation}
where the subindices $r$ and $l$ denote right and left moving
contributions, $E_r$ and $E_l$ are the corresponding amplitudes which
we take as slowly varying functions of $\vec r$, and
$\hat e_r$ and $\hat e_l$ the polarizations within the $x-y$ plane. To
ensure that the field is divergence-less, an additional field contribution
along $z$ has to be added to (\ref{E}),
\begin{equation}\label{DE}
\Delta \vec E(\vec r, t) = \left(\hat e_r\cdot\nabla E_r e^{i(q z-\omega
t)}-\hat e_l\cdot\nabla E_l e^{-i(qz+\omega t)}\right)i \hat z/q.
\end{equation}
Expressions similar to (\ref{E}) and (\ref{DE}) may also be written for
the magnetic field.
The torque
$\tau_z$ over medium 2 may be obtained by integrating the angular
momentum flux $\mathbf M=\vec r\times \mathbf T$ over a surface that
surrounds it, where $\mathbf T$ is the 
electromagnetic stress tensor. Thus,
\begin{equation}\label{tau}
\tau_z=-\frac{1}{8\pi}\mbox{Re}\int da\, [(\vec r\times \vec
E^*)_z E_z+(\vec r\times \vec B^*)_z B_z],
\end{equation}
where the integral is over the cross section of the beam. Notice that
had we started 
our calculation with an infinitely extended plane wave, $E_z$ and
$B_z$ would have been zero, but the integral in Eq. (\ref{tau}) would have
been over an infinitely extended surface, yielding an ill defined
result. On the other hand, starting from Eq. (\ref{tau}) we can take
the limit of a plane wave, obtaining well defined
expressions. The electromagnetic torque may be considered an edge effect that
survives in the plane wave limit. 
Substituting (\ref{E}) and 
(\ref{DE}) and similar expressions for the magnetic field $\vec B$,
and after some manipulation, we obtain 
\begin{equation}\label{tau1}
\tau_z=-\frac{A}{8\pi q^2}\mbox{Re}(\vec E\times\partial_z\vec E^*)_z,
\end{equation}
where $A\to\infty$ is the cross sectional area of the wavefront.
It can easily be checked that Eq. (\ref{tau1}) is consistent with the
quantum mechanical view that each photon of energy $\hbar\omega$
carries an angular momentum $\pm \hbar$ along $\pm z$ with speed $c$,
according to its helicity. 

Now we consider one normal mode  of the fictitious system with
amplitude $\mathcal E_0$ and frequency $\omega$, $\vec E=\mathcal
E_0\vec \phi(z) e^{-i\omega t}$, where 
\begin{equation}\label{phi}
\vec \phi(z)=\vec C^\Lambda e^{iqz}+\vec D^\Lambda
e^{-iqz},\quad(\Lambda=I,II,III) 
\end{equation}
is a spinorial normalized {\em wavefunction} with components
$\phi_\mu$ ($\mu=x,y$), $\vec C^\Lambda$,
$\vec D^\Lambda$ are distinct coefficients within each region
$\Lambda$ and $q=\omega/c$. In the limit $L_I,L_{III}\to\infty$, the
electromagnetic energy
$
U= [ L_I( ||C^I||^2 + ||D^I||^2) + L_{III}
  (||C^{III}||^2 + ||D^{III}||^2)] |\mathcal E_0|^2 A/8\pi
$
and the normalization condition
$
1= [ L_I( ||C^I||^2 +
  ||D^I||^2) + L_{III} (||C^{III}||^2 + ||D^{III}||^2)]
$
are dominated by the large fictitious regions I and III, so we may
identify $U=A|\mathcal E_0|^2/8\pi$ and solve for the amplitude
$|\mathcal E_0|^2=8\pi f_\omega \hbar\omega/A$ in terms of the
equilibrium photon occupation number
$f_\omega=\coth(\beta\hbar\omega/2)/2$ at temperature $k_B
T=1/\beta$. Thus, the contribution  to the torque (\ref{tau}) of one mode 
may be written as
\begin{equation}\label{tau2}
  \tau_z=-\frac{\hbar c}{2q} f_\omega[\phi_x\partial_z\phi^*_y -
    \phi_y\partial_z\phi^*_x + (\partial_z
    \phi_y)\phi^*_x-(\partial_z\phi_x)\phi^*_y].
\end{equation}
Now we label each mode by an index $n$ and sum (\ref{tau2}) over $n$
to obtain the total torque
\begin{equation}\label{total}
\fl
  \tau_z=-\hbar c\int dq  f_{qc}\sum_n
  \delta(q^2-q_n^2)
	\times [\phi_{nx}\partial_z\phi^*_{ny} - 
    \phi_{ny}\partial_z\phi^*_{nx} + (\partial_z
    \phi_{ny})\phi^*_{nx}-(\partial_z\phi_{nx})\phi^*_{ny}],
\end{equation}
where we introduced the $q$ integration and the Dirac's $\delta$ in
order to write the result in terms of the Green's function
$G_{\mu\nu}(z,z') = \sum_n \phi_{n\mu}(z) \phi_{n\nu}^*(z') /
(q^2+i\eta-q_n^2)$,
\begin{equation}\label{tauvsG}
	\tau_z=\frac{\hbar c}{\pi}\int_0^\infty dq\, f_{qc}
  (\partial_{z'}-\partial_z)[A_{xy}(z,z')-A_{yx}(z,z')]_{z'\to z},
\end{equation}
where $A_{\mu\nu}(z,z')=[G_{\mu\nu}(z,z')-G_{\nu\mu}(z',z)]/2i$
is the anti-Hermitian part of $G_{\mu\nu}$. Here, $\eta$ is a positive
infinitesimal and we employed the identity $\pi
\delta(x)=-\mbox{Im}(x+i\eta)^{-1}$. 
The Green's function may be evaluated by solving 
$
(\partial_z^2+q^2+i\eta)G_{\mu\nu}(z,z')=\delta(z-z')\delta_{\mu\nu},
$
subject to the appropriate boundary conditions. We write the solution
in terms of the two homogeneous solutions $\vec u_{\lambda}(z)$
and $\vec v_{\lambda}(z)$ ($\lambda=1,2$) 
that satisfy the boundary conditions on the right and the left side
of the system respectively,
\begin{eqnarray}\label{Gvsu}
  \mathbf G(z,z') &=& \mathbf u(z) [\mathbf u'(z)-\mathbf
  v'(z') \mathbf v^{-1}(z') \mathbf u(z')]^{-1} \theta(z-z')
\nonumber \\
&&
-\mathbf v(z) [\mathbf v'(z')-\mathbf
  u'(z') \mathbf u^{-1}(z') \mathbf v(z')]^{-1} \theta(z'-z),
\end{eqnarray}
where $\mathbf G(z,z')$ is a matrix with elements
$G_{\mu\nu}(z,z')$, $\mathbf u(z)$ and $\mathbf v(z)$ are matrices
with columns $\vec u_\lambda(z)$ and $\vec v_\lambda(z)$ respectively,
i.e., with 
matrix elements 
$u_{\mu\lambda}(z)$ and $v_{\mu\lambda}(z)$,
$\mathbf u'$ and $\mathbf v'$ denote their derivatives
with respect to their argument, and $\theta$ denotes the Heaviside
unit step function. 

For the case of uniaxial or orthorhombic slabs with normal-incidence
reflection  amplitudes $r_{x_a}$ and $r_{y_a}$ in the principal axes
$x_a$ and $y_a$ of the $a$-th slab we may write 
\begin{equation}\label{uzvz}
  \mathbf u(z)=\mathbf R\cdot \mathbf u_0(z),\quad
  \mathbf v(z)=\mathbf R^T\cdot \mathbf v_0(z),
\end{equation}
where $\mathbf R$ and $\mathbf R^T$ are rotation matrices by angles
$\gamma/2$ and $-\gamma/2$ respectively
and
\begin{equation}\label{u0}
  \mathbf u_0(z) = \left(
  \begin{array}{cc}
    1&0\\0&1
  \end{array}\right) e^{i q(z-z_2)}+
  \left(
  \begin{array}{cc}
    r_{x_2}&0\\0&r_{y_2}
  \end{array}\right) e^{-i q(z-z_2)},
\end{equation}
\begin{equation}\label{v0}
  \mathbf v_0(z) = \left(
  \begin{array}{cc}
    1&0\\0&1
  \end{array}\right) e^{-i q (z-z_1)}+
  \left(
  \begin{array}{cc}
    r_{x_1}&0\\0&r_{y_1}
  \end{array}\right) e^{i q (z-z_1)},
\end{equation}
are the solutions at the right and left sides of $\mathcal V'$
referred to the principal axes of the corresponding slab, which we
assumed to be rotated by an angle $\gamma$ with respect to those of the
opposite slab. Notice that $\mathbf R$ and $\mathbf R^T$ act only on
the first index $\mu$ of $\mathbf u$ and $\mathbf v$.

Substituting Eqs. (\ref{uzvz})-(\ref{v0}) in Eqs. (\ref{Gvsu}) and
(\ref{tauvsG}) we obtain a simple expression
\begin{equation}\label{taufin}
\fl    \tau_z=-\frac{\hbar c}{2\pi}\int_0^\infty d\kappa\,
    \frac{\Delta r_1 \Delta r_2 \sin 2\gamma\,e^{-2\kappa L}}
  {
      \begin{array}{l}
	\Delta r_1 \Delta r_2\sin^2 \gamma e^{-2\kappa L}
	+(1-r_{1x} r_{2x} e^{-2\kappa L})
	(1-r_{1y}r_{2y} e^{-2\kappa L})
      \end{array}
  },
\end{equation}
where we also took the zero temperature limit $f_{qc}=1/2$ and we
deformed the $q$ integration path from the positive real axis
toward the positive imaginary axis, as is usual. Here, $\kappa=q/i$ and we
defined the anisotropy $\Delta r_a=r_{x_a}-r_{y_a}$ of the $a$-th
slab. Eq. (\ref{taufin}) is the main result of the present paper. We
have verified that it is equivalent to the result of Ref. \cite{12},
although in a much more compact form.

\section{Results}

\begin{figure}
\centering{\input{perf}}
\caption{\label{fperf}Torque in 1D at $T=0$ between perfect mirrors
covered by perfect polarizers as a function of the angle $\gamma$
between corresponding principal directions.}
\end{figure}
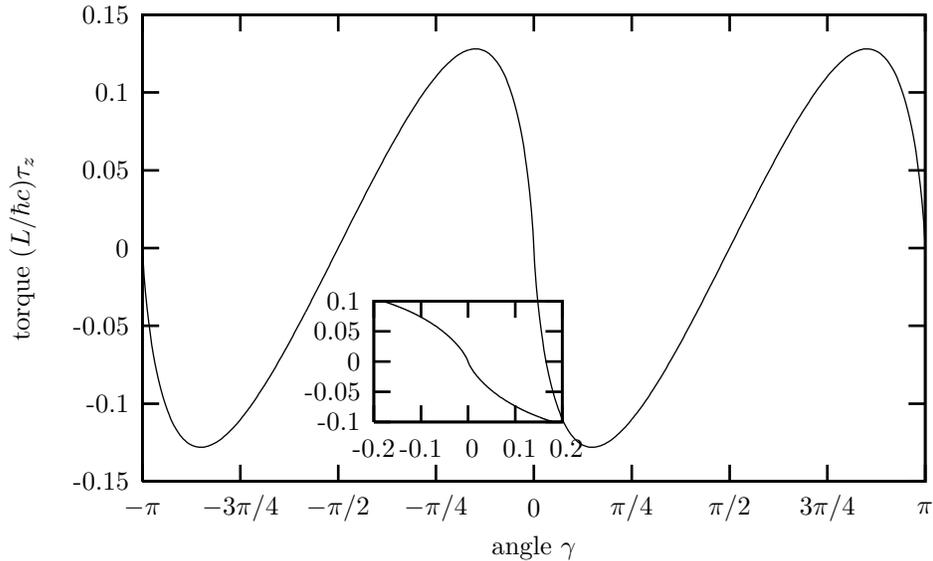
As a first application of our result (\ref{taufin}) we calculate the
torque between two ideal perfectly reflecting mirrors covered by ideal
perfectly absorbing polarizers, that is, we take $r_{x_1}=r_{x_2}=\pm
1$, $r_{y_1}=r_{y_2}=0$. In this case, Eq. (\ref{taufin}) may be
integrated analytically, yielding
\begin{equation}\label{perf}
  \tau_z= \frac{\hbar c}{2\pi  L} \tan\gamma  \,\log \sin^2 \gamma.
\end{equation}
Notice that the torque decays as $1/L$, in analogy to the
$1/L^2$ decay of the Casimir force between perfect mirrors in 1D and
the corresponding $1/L^4$ decay in 3D. In Fig. \ref{fperf} we show the
torque as a function of the angle. As could have been expected, it is
a periodic function of $\gamma$ with period $\pi$. It is null when
both polarizers are aligned, $\gamma=0$, 
corresponding to a stable equilibrium orientation, and when they are
orthogonal to each other, $\gamma=\pm \pi/2$, corresponding to
unstable equilibrium. The inset shows that the slope of $\tau_z(\gamma)$ is
singular at the stable equilibrium point. We remark that the torque is
not simply proportional to $\sin 2\gamma$ and therefore its extreme
values are not at $\gamma=\pm\pi/4$. However, it is odd-symmetric
around $\gamma=0$. The maximum torque may be estimated as $0.1\hbar
c/L$; for example, at $L=10$nm it is about
$10^{-19}$Nm. For dimensional reasons, in a full 3D
calculation our result above would have to be scaled by $A/L^2$
multiplied by some dimensionless factor. 

\begin{figure}
\centering{\input{imperf}}
\caption{\label{imperf}Torque between two lossy mirrors with reflection
amplitudes $|r|=0.6, 0.7, 0.8,  0.9,  1.0$ covered by
perfect polarizers as a function of the angle $\gamma$ between
corresponding principal directions.}
\end{figure}
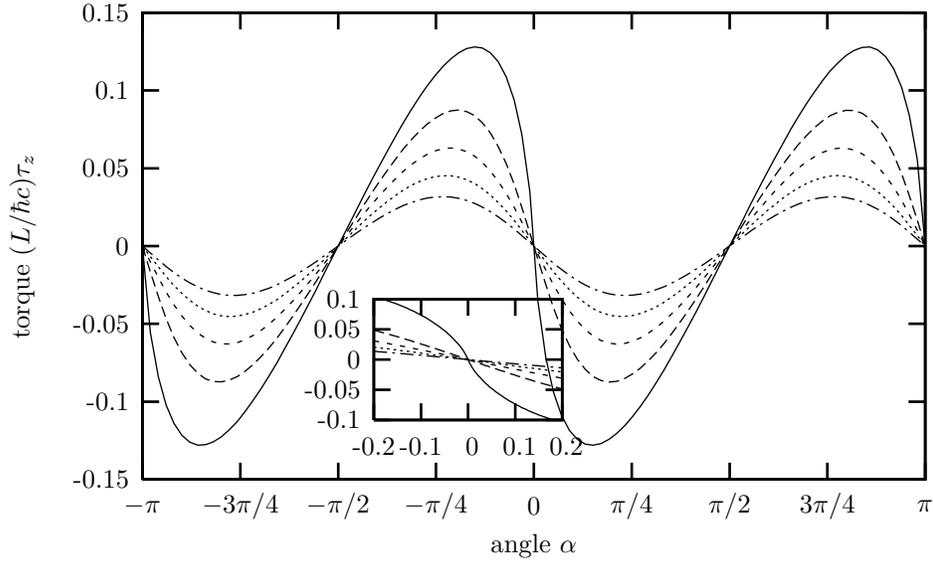
If we replace the perfect mirrors above by lossy mirrors with
reflection amplitude $r$, we can again obtain an analytical expression
\begin{equation}
  \tau_z=\frac{\hbar c\tan\gamma}{2 \pi L}
  \log  (1-|r|^2\cos^2 \gamma),
\end{equation}
Fig. \ref{imperf} shows that as $|r|$ diminishes $\tau_z$ is reduced
and becomes closer to a simple sinusoidal function
$\tau_z\approx-\hbar c |r|^2 
\sin2\gamma/4\pi L$. The 
inset shows that the singularity at $\gamma=0$ disappears when
$|r|<1$. 

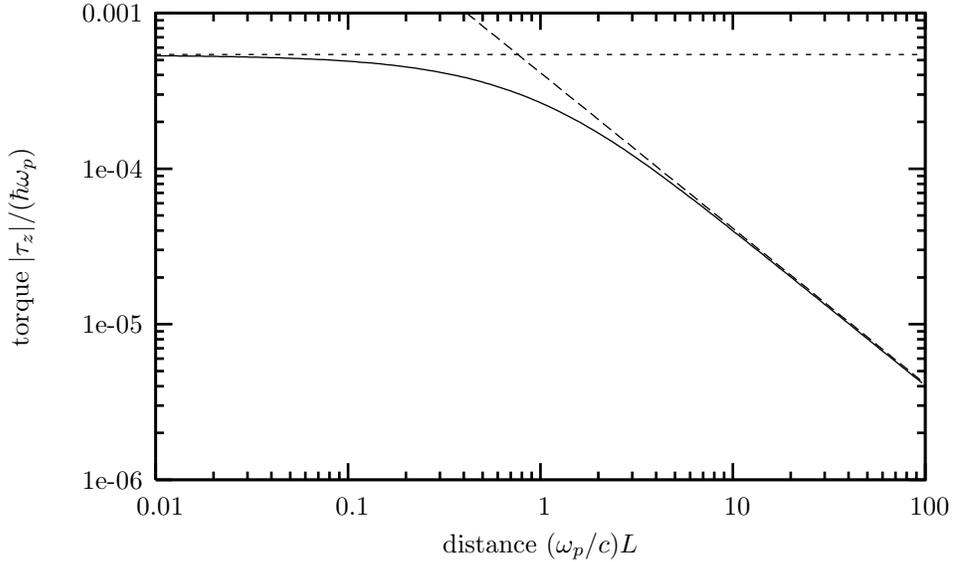
\begin{figure}
\centering{\input{dicroic}}
\caption{\label{dicroic}Torque between two identical dichroic mirrors
as a function of their separation for a fixed angle $\gamma=\pi/4$
between their principal directions. The resonance parameters are
$\omega_{x_1 p}=\omega_{y_1 p}=\omega_{x_2 p}=\omega_{y_2
p}=\omega_p$, $\omega_{x_1}=\omega_{x_2}=\omega_p$,
$\omega_{y_1}=\omega_{y_2}=\sqrt2\omega_p$.}
\end{figure}
In Fig. \ref{dicroic} we illustrate the torque between two identical
dichroic mirrors relatively rotated by $\gamma=\pi/4$ as a function of
separation. Each mirror is
characterized by a Lorentzian dielectric tensor with principal components
\begin{equation}\label{eps}
\epsilon_{\mu_a}(\omega)=1+\frac{\omega_{\mu_a p}^2}{\omega_{\mu_a}^2-\omega^2-i\omega/\tau_{\mu_a}},
\end{equation}
where $\omega_{\mu_a}$ is the frequency, $\tau_{\mu_a}$ the lifetime
and $\omega_{\mu_a p}$ the intensity of a resonance of the $a$-th slab
corresponding to polarizing field along the $\mu$-th principal
axis. In this case the characteristic frequency $\omega_p$
defines a characteristic lengthscale $c/\omega_p$. For
separations much smaller than this distance the calculation is essentially
non-retarded and the 1D torque becomes constant, proportional to $\hbar
\omega_p$. On the other hand, for larger separations  we reach the
retarded regime and the torque decays as $\hbar c/L$, as for the ideal
case.

\section{Conclusions}

Use of the scattering approach allowed us to obtain a
simple expressions for the Casimir torque between anisotropic media in
terms of their optical coefficients. Thus our results are applicable
to arbitrary anisotropic materials and not only to semiinfinite,
local, homogeneous ones. For instance, they may be readily applied to
free standing or supported anisotropic films and to heterogeneous systems.
We obtained analytical
expressions for ideal systems which are the anisotropic counterparts
to the ideal Casimir mirrors, and numerical results
covering both the retarded and non-retarded regions for realistic
dichroic systems with dispersive response functions. Our formalism has
also permitted calculations of the 
torque between dissimilar materials, suggesting procedures to
manipulate it, and it has been generalized to the full 3D
case \cite{unpub}. 

\ack
We acknowledge useful discussions with C. Villarreal and R. Esquivel.
This work was partially supported by DGAPA-UNAM under grant IN117402.

\end{document}

%% file: perf.tex
\begingroup%
  \makeatletter%
  \newcommand{\GNUPLOTspecial}{%
    \@sanitize\catcode`\%=14\relax\special}%
  \setlength{\unitlength}{0.1bp}%
\begin{picture}(3600,2160)(0,0)%
{\GNUPLOTspecial{"
/gnudict 256 dict def
gnudict begin
/Color false def
/Solid false def
/gnulinewidth 5.000 def
/userlinewidth gnulinewidth def
/vshift -33 def
/dl {10.0 mul} def
/hpt_ 31.5 def
/vpt_ 31.5 def
/hpt hpt_ def
/vpt vpt_ def
/Rounded false def
/M {moveto} bind def
/L {lineto} bind def
/R {rmoveto} bind def
/V {rlineto} bind def
/N {newpath moveto} bind def
/C {setrgbcolor} bind def
/f {rlineto fill} bind def
/vpt2 vpt 2 mul def
/hpt2 hpt 2 mul def
/Lshow { currentpoint stroke M
  0 vshift R show } def
/Rshow { currentpoint stroke M
  dup stringwidth pop neg vshift R show } def
/Cshow { currentpoint stroke M
  dup stringwidth pop -2 div vshift R show } def
/UP { dup vpt_ mul /vpt exch def hpt_ mul /hpt exch def
  /hpt2 hpt 2 mul def /vpt2 vpt 2 mul def } def
/DL { Color {setrgbcolor Solid {pop []} if 0 setdash }
 {pop pop pop 0 setgray Solid {pop []} if 0 setdash} ifelse } def
/BL { stroke userlinewidth 2 mul setlinewidth
      Rounded { 1 setlinejoin 1 setlinecap } if } def
/AL { stroke userlinewidth 2 div setlinewidth
      Rounded { 1 setlinejoin 1 setlinecap } if } def
/UL { dup gnulinewidth mul /userlinewidth exch def
      dup 1 lt {pop 1} if 10 mul /udl exch def } def
/PL { stroke userlinewidth setlinewidth
      Rounded { 1 setlinejoin 1 setlinecap } if } def
/LTw { PL [] 1 setgray } def
/LTb { BL [] 0 0 0 DL } def
/LTa { AL [1 udl mul 2 udl mul] 0 setdash 0 0 0 setrgbcolor } def
/LT0 { PL [] 1 0 0 DL } def
/LT1 { PL [4 dl 2 dl] 0 1 0 DL } def
/LT2 { PL [2 dl 3 dl] 0 0 1 DL } def
/LT3 { PL [1 dl 1.5 dl] 1 0 1 DL } def
/LT4 { PL [5 dl 2 dl 1 dl 2 dl] 0 1 1 DL } def
/LT5 { PL [4 dl 3 dl 1 dl 3 dl] 1 1 0 DL } def
/LT6 { PL [2 dl 2 dl 2 dl 4 dl] 0 0 0 DL } def
/LT7 { PL [2 dl 2 dl 2 dl 2 dl 2 dl 4 dl] 1 0.3 0 DL } def
/LT8 { PL [2 dl 2 dl 2 dl 2 dl 2 dl 2 dl 2 dl 4 dl] 0.5 0.5 0.5 DL } def
/Pnt { stroke [] 0 setdash
   gsave 1 setlinecap M 0 0 V stroke grestore } def
/Dia { stroke [] 0 setdash 2 copy vpt add M
  hpt neg vpt neg V hpt vpt neg V
  hpt vpt V hpt neg vpt V closepath stroke
  Pnt } def
/Pls { stroke [] 0 setdash vpt sub M 0 vpt2 V
  currentpoint stroke M
  hpt neg vpt neg R hpt2 0 V stroke
  } def
/Box { stroke [] 0 setdash 2 copy exch hpt sub exch vpt add M
  0 vpt2 neg V hpt2 0 V 0 vpt2 V
  hpt2 neg 0 V closepath stroke
  Pnt } def
/Crs { stroke [] 0 setdash exch hpt sub exch vpt add M
  hpt2 vpt2 neg V currentpoint stroke M
  hpt2 neg 0 R hpt2 vpt2 V stroke } def
/TriU { stroke [] 0 setdash 2 copy vpt 1.12 mul add M
  hpt neg vpt -1.62 mul V
  hpt 2 mul 0 V
  hpt neg vpt 1.62 mul V closepath stroke
  Pnt  } def
/Star { 2 copy Pls Crs } def
/BoxF { stroke [] 0 setdash exch hpt sub exch vpt add M
  0 vpt2 neg V  hpt2 0 V  0 vpt2 V
  hpt2 neg 0 V  closepath fill } def
/TriUF { stroke [] 0 setdash vpt 1.12 mul add M
  hpt neg vpt -1.62 mul V
  hpt 2 mul 0 V
  hpt neg vpt 1.62 mul V closepath fill } def
/TriD { stroke [] 0 setdash 2 copy vpt 1.12 mul sub M
  hpt neg vpt 1.62 mul V
  hpt 2 mul 0 V
  hpt neg vpt -1.62 mul V closepath stroke
  Pnt  } def
/TriDF { stroke [] 0 setdash vpt 1.12 mul sub M
  hpt neg vpt 1.62 mul V
  hpt 2 mul 0 V
  hpt neg vpt -1.62 mul V closepath fill} def
/DiaF { stroke [] 0 setdash vpt add M
  hpt neg vpt neg V hpt vpt neg V
  hpt vpt V hpt neg vpt V closepath fill } def
/Pent { stroke [] 0 setdash 2 copy gsave
  translate 0 hpt M 4 {72 rotate 0 hpt L} repeat
  closepath stroke grestore Pnt } def
/PentF { stroke [] 0 setdash gsave
  translate 0 hpt M 4 {72 rotate 0 hpt L} repeat
  closepath fill grestore } def
/Circle { stroke [] 0 setdash 2 copy
  hpt 0 360 arc stroke Pnt } def
/CircleF { stroke [] 0 setdash hpt 0 360 arc fill } def
/C0 { BL [] 0 setdash 2 copy moveto vpt 90 450  arc } bind def
/C1 { BL [] 0 setdash 2 copy        moveto
       2 copy  vpt 0 90 arc closepath fill
               vpt 0 360 arc closepath } bind def
/C2 { BL [] 0 setdash 2 copy moveto
       2 copy  vpt 90 180 arc closepath fill
               vpt 0 360 arc closepath } bind def
/C3 { BL [] 0 setdash 2 copy moveto
       2 copy  vpt 0 180 arc closepath fill
               vpt 0 360 arc closepath } bind def
/C4 { BL [] 0 setdash 2 copy moveto
       2 copy  vpt 180 270 arc closepath fill
               vpt 0 360 arc closepath } bind def
/C5 { BL [] 0 setdash 2 copy moveto
       2 copy  vpt 0 90 arc
       2 copy moveto
       2 copy  vpt 180 270 arc closepath fill
               vpt 0 360 arc } bind def
/C6 { BL [] 0 setdash 2 copy moveto
      2 copy  vpt 90 270 arc closepath fill
              vpt 0 360 arc closepath } bind def
/C7 { BL [] 0 setdash 2 copy moveto
      2 copy  vpt 0 270 arc closepath fill
              vpt 0 360 arc closepath } bind def
/C8 { BL [] 0 setdash 2 copy moveto
      2 copy vpt 270 360 arc closepath fill
              vpt 0 360 arc closepath } bind def
/C9 { BL [] 0 setdash 2 copy moveto
      2 copy  vpt 270 450 arc closepath fill
              vpt 0 360 arc closepath } bind def
/C10 { BL [] 0 setdash 2 copy 2 copy moveto vpt 270 360 arc closepath fill
       2 copy moveto
       2 copy vpt 90 180 arc closepath fill
               vpt 0 360 arc closepath } bind def
/C11 { BL [] 0 setdash 2 copy moveto
       2 copy  vpt 0 180 arc closepath fill
       2 copy moveto
       2 copy  vpt 270 360 arc closepath fill
               vpt 0 360 arc closepath } bind def
/C12 { BL [] 0 setdash 2 copy moveto
       2 copy  vpt 180 360 arc closepath fill
               vpt 0 360 arc closepath } bind def
/C13 { BL [] 0 setdash  2 copy moveto
       2 copy  vpt 0 90 arc closepath fill
       2 copy moveto
       2 copy  vpt 180 360 arc closepath fill
               vpt 0 360 arc closepath } bind def
/C14 { BL [] 0 setdash 2 copy moveto
       2 copy  vpt 90 360 arc closepath fill
               vpt 0 360 arc } bind def
/C15 { BL [] 0 setdash 2 copy vpt 0 360 arc closepath fill
               vpt 0 360 arc closepath } bind def
/Rec   { newpath 4 2 roll moveto 1 index 0 rlineto 0 exch rlineto
       neg 0 rlineto closepath } bind def
/Square { dup Rec } bind def
/Bsquare { vpt sub exch vpt sub exch vpt2 Square } bind def
/S0 { BL [] 0 setdash 2 copy moveto 0 vpt rlineto BL Bsquare } bind def
/S1 { BL [] 0 setdash 2 copy vpt Square fill Bsquare } bind def
/S2 { BL [] 0 setdash 2 copy exch vpt sub exch vpt Square fill Bsquare } bind def
/S3 { BL [] 0 setdash 2 copy exch vpt sub exch vpt2 vpt Rec fill Bsquare } bind def
/S4 { BL [] 0 setdash 2 copy exch vpt sub exch vpt sub vpt Square fill Bsquare } bind def
/S5 { BL [] 0 setdash 2 copy 2 copy vpt Square fill
       exch vpt sub exch vpt sub vpt Square fill Bsquare } bind def
/S6 { BL [] 0 setdash 2 copy exch vpt sub exch vpt sub vpt vpt2 Rec fill Bsquare } bind def
/S7 { BL [] 0 setdash 2 copy exch vpt sub exch vpt sub vpt vpt2 Rec fill
       2 copy vpt Square fill
       Bsquare } bind def
/S8 { BL [] 0 setdash 2 copy vpt sub vpt Square fill Bsquare } bind def
/S9 { BL [] 0 setdash 2 copy vpt sub vpt vpt2 Rec fill Bsquare } bind def
/S10 { BL [] 0 setdash 2 copy vpt sub vpt Square fill 2 copy exch vpt sub exch vpt Square fill
       Bsquare } bind def
/S11 { BL [] 0 setdash 2 copy vpt sub vpt Square fill 2 copy exch vpt sub exch vpt2 vpt Rec fill
       Bsquare } bind def
/S12 { BL [] 0 setdash 2 copy exch vpt sub exch vpt sub vpt2 vpt Rec fill Bsquare } bind def
/S13 { BL [] 0 setdash 2 copy exch vpt sub exch vpt sub vpt2 vpt Rec fill
       2 copy vpt Square fill Bsquare } bind def
/S14 { BL [] 0 setdash 2 copy exch vpt sub exch vpt sub vpt2 vpt Rec fill
       2 copy exch vpt sub exch vpt Square fill Bsquare } bind def
/S15 { BL [] 0 setdash 2 copy Bsquare fill Bsquare } bind def
/D0 { gsave translate 45 rotate 0 0 S0 stroke grestore } bind def
/D1 { gsave translate 45 rotate 0 0 S1 stroke grestore } bind def
/D2 { gsave translate 45 rotate 0 0 S2 stroke grestore } bind def
/D3 { gsave translate 45 rotate 0 0 S3 stroke grestore } bind def
/D4 { gsave translate 45 rotate 0 0 S4 stroke grestore } bind def
/D5 { gsave translate 45 rotate 0 0 S5 stroke grestore } bind def
/D6 { gsave translate 45 rotate 0 0 S6 stroke grestore } bind def
/D7 { gsave translate 45 rotate 0 0 S7 stroke grestore } bind def
/D8 { gsave translate 45 rotate 0 0 S8 stroke grestore } bind def
/D9 { gsave translate 45 rotate 0 0 S9 stroke grestore } bind def
/D10 { gsave translate 45 rotate 0 0 S10 stroke grestore } bind def
/D11 { gsave translate 45 rotate 0 0 S11 stroke grestore } bind def
/D12 { gsave translate 45 rotate 0 0 S12 stroke grestore } bind def
/D13 { gsave translate 45 rotate 0 0 S13 stroke grestore } bind def
/D14 { gsave translate 45 rotate 0 0 S14 stroke grestore } bind def
/D15 { gsave translate 45 rotate 0 0 S15 stroke grestore } bind def
/DiaE { stroke [] 0 setdash vpt add M
  hpt neg vpt neg V hpt vpt neg V
  hpt vpt V hpt neg vpt V closepath stroke } def
/BoxE { stroke [] 0 setdash exch hpt sub exch vpt add M
  0 vpt2 neg V hpt2 0 V 0 vpt2 V
  hpt2 neg 0 V closepath stroke } def
/TriUE { stroke [] 0 setdash vpt 1.12 mul add M
  hpt neg vpt -1.62 mul V
  hpt 2 mul 0 V
  hpt neg vpt 1.62 mul V closepath stroke } def
/TriDE { stroke [] 0 setdash vpt 1.12 mul sub M
  hpt neg vpt 1.62 mul V
  hpt 2 mul 0 V
  hpt neg vpt -1.62 mul V closepath stroke } def
/PentE { stroke [] 0 setdash gsave
  translate 0 hpt M 4 {72 rotate 0 hpt L} repeat
  closepath stroke grestore } def
/CircE { stroke [] 0 setdash 
  hpt 0 360 arc stroke } def
/Opaque { gsave closepath 1 setgray fill grestore 0 setgray closepath } def
/DiaW { stroke [] 0 setdash vpt add M
  hpt neg vpt neg V hpt vpt neg V
  hpt vpt V hpt neg vpt V Opaque stroke } def
/BoxW { stroke [] 0 setdash exch hpt sub exch vpt add M
  0 vpt2 neg V hpt2 0 V 0 vpt2 V
  hpt2 neg 0 V Opaque stroke } def
/TriUW { stroke [] 0 setdash vpt 1.12 mul add M
  hpt neg vpt -1.62 mul V
  hpt 2 mul 0 V
  hpt neg vpt 1.62 mul V Opaque stroke } def
/TriDW { stroke [] 0 setdash vpt 1.12 mul sub M
  hpt neg vpt 1.62 mul V
  hpt 2 mul 0 V
  hpt neg vpt -1.62 mul V Opaque stroke } def
/PentW { stroke [] 0 setdash gsave
  translate 0 hpt M 4 {72 rotate 0 hpt L} repeat
  Opaque stroke grestore } def
/CircW { stroke [] 0 setdash 
  hpt 0 360 arc Opaque stroke } def
/BoxFill { gsave Rec 1 setgray fill grestore } def
/BoxColFill {
  gsave Rec
  /Fillden exch def
  currentrgbcolor
  /ColB exch def /ColG exch def /ColR exch def
  /ColR ColR Fillden mul Fillden sub 1 add def
  /ColG ColG Fillden mul Fillden sub 1 add def
  /ColB ColB Fillden mul Fillden sub 1 add def
  ColR ColG ColB setrgbcolor
  fill grestore } def
%
%
/PatternFill { gsave /PFa [ 9 2 roll ] def
    PFa 0 get PFa 2 get 2 div add PFa 1 get PFa 3 get 2 div add translate
    PFa 2 get -2 div PFa 3 get -2 div PFa 2 get PFa 3 get Rec
    gsave 1 setgray fill grestore clip
    currentlinewidth 0.5 mul setlinewidth
    /PFs PFa 2 get dup mul PFa 3 get dup mul add sqrt def
    0 0 M PFa 5 get rotate PFs -2 div dup translate
	0 1 PFs PFa 4 get div 1 add floor cvi
	{ PFa 4 get mul 0 M 0 PFs V } for
    0 PFa 6 get ne {
	0 1 PFs PFa 4 get div 1 add floor cvi
	{ PFa 4 get mul 0 2 1 roll M PFs 0 V } for
    } if
    stroke grestore } def
/Symbol-Oblique /Symbol findfont [1 0 .167 1 0 0] makefont
dup length dict begin {1 index /FID eq {pop pop} {def} ifelse} forall
currentdict end definefont pop
end
gnudict begin
gsave
0 0 translate
0.100 0.100 scale
0 setgray
newpath
1.000 UL
LTb
500 300 M
63 0 V
2887 0 R
-63 0 V
1.000 UL
LTb
500 593 M
63 0 V
2887 0 R
-63 0 V
1.000 UL
LTb
500 887 M
63 0 V
2887 0 R
-63 0 V
1.000 UL
LTb
500 1180 M
63 0 V
2887 0 R
-63 0 V
1.000 UL
LTb
500 1473 M
63 0 V
2887 0 R
-63 0 V
1.000 UL
LTb
500 1767 M
63 0 V
2887 0 R
-63 0 V
1.000 UL
LTb
500 2060 M
63 0 V
2887 0 R
-63 0 V
1.000 UL
LTb
500 300 M
0 63 V
0 1697 R
0 -63 V
1.000 UL
LTb
869 300 M
0 63 V
0 1697 R
0 -63 V
1.000 UL
LTb
1238 300 M
0 63 V
0 1697 R
0 -63 V
1.000 UL
LTb
1606 300 M
0 63 V
0 1697 R
0 -63 V
1.000 UL
LTb
1975 300 M
0 63 V
0 1697 R
0 -63 V
1.000 UL
LTb
2344 300 M
0 63 V
0 1697 R
0 -63 V
1.000 UL
LTb
2713 300 M
0 63 V
0 1697 R
0 -63 V
1.000 UL
LTb
3081 300 M
0 63 V
0 1697 R
0 -63 V
1.000 UL
LTb
3450 300 M
0 63 V
0 1697 R
0 -63 V
1.000 UL
LTb
1.000 UL
LTb
500 300 M
2950 0 V
0 1760 V
-2950 0 V
500 300 L
LTb
LTb
1.000 UP
1.000 UL
LT0
500 1180 M
6 -103 V
6 -70 V
6 -58 V
6 -50 V
6 -45 V
5 -39 V
6 -36 V
6 -33 V
6 -29 V
6 -27 V
6 -25 V
6 -23 V
6 -22 V
6 -19 V
6 -18 V
6 -17 V
6 -15 V
5 -14 V
6 -13 V
6 -12 V
6 -10 V
6 -10 V
6 -9 V
6 -9 V
6 -7 V
6 -7 V
6 -5 V
6 -6 V
5 -4 V
6 -4 V
6 -3 V
6 -3 V
6 -2 V
6 -2 V
6 -1 V
6 -1 V
6 0 V
6 0 V
6 0 V
5 1 V
6 2 V
6 1 V
6 3 V
6 2 V
6 3 V
6 3 V
6 3 V
6 4 V
6 4 V
6 4 V
6 5 V
5 5 V
6 5 V
6 5 V
6 6 V
6 6 V
6 6 V
6 6 V
6 7 V
6 6 V
6 7 V
6 7 V
5 7 V
6 8 V
6 7 V
6 8 V
6 8 V
6 8 V
6 8 V
6 9 V
6 8 V
6 9 V
6 9 V
5 9 V
6 9 V
6 9 V
6 9 V
6 9 V
6 10 V
6 10 V
6 9 V
6 10 V
6 10 V
6 10 V
6 10 V
5 10 V
6 11 V
6 10 V
6 10 V
6 11 V
6 11 V
6 10 V
6 11 V
6 11 V
6 11 V
6 11 V
5 11 V
6 11 V
6 11 V
6 11 V
6 11 V
6 11 V
6 12 V
6 11 V
stroke
1115 939 M
6 11 V
6 12 V
6 11 V
5 12 V
6 11 V
6 12 V
6 11 V
6 12 V
6 11 V
6 12 V
6 12 V
6 11 V
6 12 V
6 12 V
6 11 V
5 12 V
6 12 V
6 12 V
6 11 V
6 12 V
6 12 V
6 12 V
6 11 V
6 12 V
6 12 V
6 12 V
5 11 V
6 12 V
6 12 V
6 12 V
6 11 V
6 12 V
6 11 V
6 12 V
6 12 V
6 11 V
6 12 V
5 11 V
6 12 V
6 11 V
6 11 V
6 12 V
6 11 V
6 11 V
6 12 V
6 11 V
6 11 V
6 11 V
6 11 V
5 11 V
6 11 V
6 11 V
6 10 V
6 11 V
6 11 V
6 10 V
6 11 V
6 10 V
6 10 V
6 11 V
5 10 V
6 10 V
6 10 V
6 10 V
6 10 V
6 9 V
6 10 V
6 9 V
6 10 V
6 9 V
6 9 V
5 9 V
6 9 V
6 8 V
6 9 V
6 9 V
6 8 V
6 8 V
6 8 V
6 8 V
6 7 V
6 8 V
6 7 V
5 7 V
6 7 V
6 7 V
6 7 V
6 6 V
6 6 V
6 6 V
6 6 V
6 5 V
6 6 V
6 5 V
5 4 V
6 5 V
6 4 V
6 4 V
6 3 V
6 4 V
6 3 V
6 2 V
6 3 V
6 1 V
stroke
1730 1927 M
6 2 V
5 1 V
6 1 V
6 0 V
6 0 V
6 0 V
6 -1 V
6 -2 V
6 -2 V
6 -2 V
6 -3 V
6 -4 V
6 -4 V
5 -5 V
6 -5 V
6 -7 V
6 -7 V
6 -7 V
6 -9 V
6 -9 V
6 -11 V
6 -11 V
6 -12 V
6 -14 V
5 -15 V
6 -15 V
6 -18 V
6 -18 V
6 -21 V
6 -22 V
6 -24 V
6 -26 V
6 -28 V
6 -31 V
6 -34 V
5 -38 V
6 -42 V
6 -47 V
6 -54 V
6 -63 V
6 -80 V
6 -120 V
6 -80 V
6 -63 V
6 -54 V
6 -47 V
6 -42 V
5 -38 V
6 -34 V
6 -31 V
6 -28 V
6 -26 V
6 -24 V
6 -22 V
6 -21 V
6 -18 V
6 -18 V
6 -15 V
5 -15 V
6 -14 V
6 -12 V
6 -11 V
6 -11 V
6 -9 V
6 -9 V
6 -7 V
6 -7 V
6 -7 V
6 -5 V
5 -5 V
6 -4 V
6 -4 V
6 -3 V
6 -2 V
6 -2 V
6 -2 V
6 -1 V
6 0 V
6 0 V
6 0 V
6 1 V
5 1 V
6 2 V
6 1 V
6 3 V
6 2 V
6 3 V
6 4 V
6 3 V
6 4 V
6 4 V
6 5 V
5 4 V
6 5 V
6 6 V
6 5 V
6 6 V
6 6 V
6 6 V
6 6 V
6 7 V
6 7 V
6 7 V
5 7 V
stroke
2344 534 M
6 7 V
6 8 V
6 7 V
6 8 V
6 8 V
6 8 V
6 8 V
6 9 V
6 9 V
6 8 V
6 9 V
5 9 V
6 9 V
6 9 V
6 10 V
6 9 V
6 10 V
6 9 V
6 10 V
6 10 V
6 10 V
6 10 V
5 10 V
6 11 V
6 10 V
6 10 V
6 11 V
6 10 V
6 11 V
6 11 V
6 10 V
6 11 V
6 11 V
5 11 V
6 11 V
6 11 V
6 11 V
6 11 V
6 12 V
6 11 V
6 11 V
6 12 V
6 11 V
6 11 V
6 12 V
5 11 V
6 12 V
6 11 V
6 12 V
6 12 V
6 11 V
6 12 V
6 11 V
6 12 V
6 12 V
6 12 V
5 11 V
6 12 V
6 12 V
6 12 V
6 11 V
6 12 V
6 12 V
6 12 V
6 11 V
6 12 V
6 12 V
5 12 V
6 11 V
6 12 V
6 12 V
6 11 V
6 12 V
6 12 V
6 11 V
6 12 V
6 11 V
6 12 V
6 11 V
5 12 V
6 11 V
6 12 V
6 11 V
6 11 V
6 12 V
6 11 V
6 11 V
6 11 V
6 11 V
6 11 V
5 11 V
6 11 V
6 11 V
6 11 V
6 11 V
6 10 V
6 11 V
6 11 V
6 10 V
6 10 V
6 11 V
5 10 V
6 10 V
6 10 V
stroke
2959 1647 M
6 10 V
6 10 V
6 9 V
6 10 V
6 10 V
6 9 V
6 9 V
6 9 V
6 9 V
5 9 V
6 9 V
6 9 V
6 8 V
6 9 V
6 8 V
6 8 V
6 8 V
6 8 V
6 7 V
6 8 V
5 7 V
6 7 V
6 7 V
6 6 V
6 7 V
6 6 V
6 6 V
6 6 V
6 6 V
6 5 V
6 5 V
5 5 V
6 5 V
6 4 V
6 4 V
6 4 V
6 3 V
6 3 V
6 3 V
6 2 V
6 3 V
6 1 V
6 2 V
5 1 V
6 0 V
6 0 V
6 0 V
6 -1 V
6 -1 V
6 -2 V
6 -2 V
6 -3 V
6 -3 V
6 -4 V
5 -4 V
6 -6 V
6 -5 V
6 -7 V
6 -7 V
6 -9 V
6 -9 V
6 -10 V
6 -10 V
6 -12 V
6 -13 V
5 -14 V
6 -15 V
6 -17 V
6 -18 V
6 -19 V
6 -22 V
6 -23 V
6 -25 V
6 -27 V
6 -29 V
6 -33 V
6 -36 V
5 -39 V
6 -45 V
6 -50 V
6 -58 V
6 -70 V
6 -103 V
1.000 UL
LTb
500 300 M
2950 0 V
0 1760 V
-2950 0 V
500 300 L
1.000 UP
1.000 UL
LTb
1372 524 M
63 0 V
647 0 R
-63 0 V
1.000 UL
LTb
1372 638 M
63 0 V
647 0 R
-63 0 V
1.000 UL
LTb
1372 752 M
63 0 V
647 0 R
-63 0 V
1.000 UL
LTb
1372 866 M
63 0 V
647 0 R
-63 0 V
1.000 UL
LTb
1372 980 M
63 0 V
647 0 R
-63 0 V
1.000 UL
LTb
1372 524 M
0 63 V
0 393 R
0 -63 V
1.000 UL
LTb
1550 524 M
0 63 V
0 393 R
0 -63 V
1.000 UL
LTb
1727 524 M
0 63 V
0 393 R
0 -63 V
1.000 UL
LTb
1905 524 M
0 63 V
0 393 R
0 -63 V
1.000 UL
LTb
2082 524 M
0 63 V
0 393 R
0 -63 V
1.000 UL
LTb
1.000 UL
LTb
1372 524 M
710 0 V
0 456 V
-710 0 V
0 -456 V
1.000 UP
1.000 UL
LT0
1405 980 M
1 0 V
2 -1 V
1 0 V
1 -1 V
2 0 V
1 -1 V
2 0 V
1 -1 V
2 0 V
1 -1 V
1 0 V
2 0 V
1 -1 V
2 0 V
1 -1 V
1 0 V
2 -1 V
1 0 V
2 -1 V
1 0 V
2 -1 V
1 0 V
1 -1 V
2 0 V
1 -1 V
2 0 V
1 -1 V
2 0 V
1 -1 V
1 0 V
2 -1 V
1 -1 V
2 0 V
1 -1 V
2 0 V
1 -1 V
1 0 V
2 -1 V
1 0 V
2 -1 V
1 0 V
1 -1 V
2 -1 V
1 0 V
2 -1 V
1 0 V
2 -1 V
1 0 V
1 -1 V
2 -1 V
1 0 V
2 -1 V
1 0 V
2 -1 V
1 0 V
1 -1 V
2 -1 V
1 0 V
2 -1 V
1 0 V
2 -1 V
1 -1 V
1 0 V
2 -1 V
1 -1 V
2 0 V
1 -1 V
1 0 V
2 -1 V
1 -1 V
2 0 V
1 -1 V
2 -1 V
1 0 V
1 -1 V
2 -1 V
1 0 V
2 -1 V
1 -1 V
2 0 V
1 -1 V
1 -1 V
2 0 V
1 -1 V
2 -1 V
1 -1 V
2 0 V
1 -1 V
1 -1 V
2 0 V
1 -1 V
2 -1 V
1 -1 V
1 0 V
2 -1 V
1 -1 V
2 -1 V
1 0 V
2 -1 V
1 -1 V
1 -1 V
2 0 V
1 -1 V
2 -1 V
stroke
1553 918 M
1 -1 V
2 -1 V
1 0 V
1 -1 V
2 -1 V
1 -1 V
2 -1 V
1 0 V
2 -1 V
1 -1 V
1 -1 V
2 -1 V
1 -1 V
2 0 V
1 -1 V
1 -1 V
2 -1 V
1 -1 V
2 -1 V
1 -1 V
2 0 V
1 -1 V
1 -1 V
2 -1 V
1 -1 V
2 -1 V
1 -1 V
2 -1 V
1 -1 V
1 -1 V
2 -1 V
1 -1 V
2 -1 V
1 -1 V
2 0 V
1 -1 V
1 -1 V
2 -1 V
1 -1 V
2 -1 V
1 -1 V
1 -1 V
2 -1 V
1 -1 V
2 -2 V
1 -1 V
2 -1 V
1 -1 V
1 -1 V
2 -1 V
1 -1 V
2 -1 V
1 -1 V
2 -1 V
1 -1 V
1 -1 V
2 -2 V
1 -1 V
2 -1 V
1 -1 V
1 -1 V
2 -1 V
1 -2 V
2 -1 V
1 -1 V
2 -1 V
1 -1 V
1 -2 V
2 -1 V
1 -1 V
2 -1 V
1 -2 V
2 -1 V
1 -1 V
1 -2 V
2 -1 V
1 -1 V
2 -2 V
1 -1 V
2 -1 V
1 -2 V
1 -1 V
2 -2 V
1 -1 V
2 -2 V
1 -1 V
1 -1 V
2 -2 V
1 -1 V
2 -2 V
1 -2 V
2 -1 V
1 -2 V
1 -1 V
2 -2 V
1 -2 V
2 -1 V
1 -2 V
2 -2 V
1 -1 V
1 -2 V
2 -2 V
1 -2 V
2 -2 V
stroke
1701 797 M
1 -2 V
2 -1 V
1 -2 V
1 -2 V
2 -2 V
1 -3 V
2 -2 V
1 -2 V
1 -2 V
2 -2 V
1 -3 V
2 -2 V
1 -3 V
2 -2 V
1 -3 V
1 -3 V
2 -3 V
1 -4 V
2 -4 V
1 -4 V
2 -3 V
1 -3 V
1 -3 V
2 -2 V
1 -3 V
2 -2 V
1 -3 V
2 -2 V
1 -2 V
1 -2 V
2 -2 V
1 -3 V
2 -2 V
1 -2 V
1 -2 V
2 -1 V
1 -2 V
2 -2 V
1 -2 V
2 -2 V
1 -2 V
1 -1 V
2 -2 V
1 -2 V
2 -1 V
1 -2 V
2 -2 V
1 -1 V
1 -2 V
2 -1 V
1 -2 V
2 -2 V
1 -1 V
2 -2 V
1 -1 V
1 -1 V
2 -2 V
1 -1 V
2 -2 V
1 -1 V
1 -2 V
2 -1 V
1 -1 V
2 -2 V
1 -1 V
2 -1 V
1 -2 V
1 -1 V
2 -1 V
1 -2 V
2 -1 V
1 -1 V
2 -1 V
1 -2 V
1 -1 V
2 -1 V
1 -1 V
2 -1 V
1 -2 V
2 -1 V
1 -1 V
1 -1 V
2 -1 V
1 -1 V
2 -2 V
1 -1 V
1 -1 V
2 -1 V
1 -1 V
2 -1 V
1 -1 V
2 -1 V
1 -1 V
1 -1 V
2 -1 V
1 -1 V
2 -2 V
1 -1 V
2 -1 V
1 -1 V
1 -1 V
2 -1 V
1 -1 V
2 -1 V
stroke
1849 618 M
1 -1 V
1 -1 V
2 0 V
1 -1 V
2 -1 V
1 -1 V
2 -1 V
1 -1 V
1 -1 V
2 -1 V
1 -1 V
2 -1 V
1 -1 V
2 -1 V
1 -1 V
1 -1 V
2 0 V
1 -1 V
2 -1 V
1 -1 V
2 -1 V
1 -1 V
1 -1 V
2 0 V
1 -1 V
2 -1 V
1 -1 V
1 -1 V
2 -1 V
1 0 V
2 -1 V
1 -1 V
2 -1 V
1 -1 V
1 0 V
2 -1 V
1 -1 V
2 -1 V
1 -1 V
2 0 V
1 -1 V
1 -1 V
2 -1 V
1 0 V
2 -1 V
1 -1 V
2 -1 V
1 0 V
1 -1 V
2 -1 V
1 -1 V
2 0 V
1 -1 V
1 -1 V
2 0 V
1 -1 V
2 -1 V
1 -1 V
2 0 V
1 -1 V
1 -1 V
2 0 V
1 -1 V
2 -1 V
1 0 V
2 -1 V
1 -1 V
1 0 V
2 -1 V
1 -1 V
2 0 V
1 -1 V
2 -1 V
1 0 V
1 -1 V
2 0 V
1 -1 V
2 -1 V
1 0 V
1 -1 V
2 -1 V
1 0 V
2 -1 V
1 0 V
2 -1 V
1 -1 V
1 0 V
2 -1 V
1 0 V
2 -1 V
1 0 V
2 -1 V
1 -1 V
1 0 V
2 -1 V
1 0 V
2 -1 V
1 0 V
2 -1 V
1 -1 V
1 0 V
2 -1 V
1 0 V
2 -1 V
stroke
1997 542 M
1 0 V
1 -1 V
2 0 V
1 -1 V
2 0 V
1 -1 V
2 -1 V
1 0 V
1 -1 V
2 0 V
1 -1 V
2 0 V
1 -1 V
2 0 V
1 -1 V
1 0 V
2 -1 V
1 0 V
2 -1 V
1 0 V
2 -1 V
1 0 V
1 -1 V
2 0 V
1 -1 V
2 0 V
1 0 V
1 -1 V
2 0 V
1 -1 V
2 0 V
1 -1 V
2 0 V
1 -1 V
1 0 V
2 -1 V
1 0 V
1.000 UL
LTb
1372 524 M
710 0 V
0 456 V
-710 0 V
0 -456 V
1.000 UP
stroke
grestore
end
showpage
}}%
\put(2082,424){\makebox(0,0){ 0.2}}%
\put(1905,424){\makebox(0,0){ 0.1}}%
\put(1727,424){\makebox(0,0){ 0}}%
\put(1550,424){\makebox(0,0){-0.1}}%
\put(1372,424){\makebox(0,0){-0.2}}%
\put(1322,980){\makebox(0,0)[r]{ 0.1}}%
\put(1322,866){\makebox(0,0)[r]{ 0.05}}%
\put(1322,752){\makebox(0,0)[r]{ 0}}%
\put(1322,638){\makebox(0,0)[r]{-0.05}}%
\put(1322,524){\makebox(0,0)[r]{-0.1}}%
\put(1975,50){\makebox(0,0){angle $\gamma$}}%
\put(100,1180){%
\special{ps: gsave currentpoint currentpoint translate
270 rotate neg exch neg exch translate}%
\makebox(0,0)[b]{\shortstack{torque $(L/\hbar c)\tau_z$}}%
\special{ps: currentpoint grestore moveto}%
}%
\put(3450,200){\makebox(0,0){$\pi$}}%
\put(3081,200){\makebox(0,0){$3\pi/4$}}%
\put(2713,200){\makebox(0,0){$\pi/2$}}%
\put(2344,200){\makebox(0,0){$\pi/4$}}%
\put(1975,200){\makebox(0,0){0}}%
\put(1606,200){\makebox(0,0){$-\pi/4$}}%
\put(1238,200){\makebox(0,0){$-\pi/2$}}%
\put(869,200){\makebox(0,0){$-3\pi/4$}}%
\put(500,200){\makebox(0,0){$-\pi$}}%
\put(450,2060){\makebox(0,0)[r]{ 0.15}}%
\put(450,1767){\makebox(0,0)[r]{ 0.1}}%
\put(450,1473){\makebox(0,0)[r]{ 0.05}}%
\put(450,1180){\makebox(0,0)[r]{ 0}}%
\put(450,887){\makebox(0,0)[r]{-0.05}}%
\put(450,593){\makebox(0,0)[r]{-0.1}}%
\put(450,300){\makebox(0,0)[r]{-0.15}}%
\end{picture}%
\endgroup
 

%% file: imperf.tex
\begingroup%
  \makeatletter%
  \newcommand{\GNUPLOTspecial}{%
    \@sanitize\catcode`\%=14\relax\special}%
  \setlength{\unitlength}{0.1bp}%
\begin{picture}(3600,2160)(0,0)%
{\GNUPLOTspecial{"
/gnudict 256 dict def
gnudict begin
/Color false def
/Solid false def
/gnulinewidth 5.000 def
/userlinewidth gnulinewidth def
/vshift -33 def
/dl {10.0 mul} def
/hpt_ 31.5 def
/vpt_ 31.5 def
/hpt hpt_ def
/vpt vpt_ def
/Rounded false def
/M {moveto} bind def
/L {lineto} bind def
/R {rmoveto} bind def
/V {rlineto} bind def
/N {newpath moveto} bind def
/C {setrgbcolor} bind def
/f {rlineto fill} bind def
/vpt2 vpt 2 mul def
/hpt2 hpt 2 mul def
/Lshow { currentpoint stroke M
  0 vshift R show } def
/Rshow { currentpoint stroke M
  dup stringwidth pop neg vshift R show } def
/Cshow { currentpoint stroke M
  dup stringwidth pop -2 div vshift R show } def
/UP { dup vpt_ mul /vpt exch def hpt_ mul /hpt exch def
  /hpt2 hpt 2 mul def /vpt2 vpt 2 mul def } def
/DL { Color {setrgbcolor Solid {pop []} if 0 setdash }
 {pop pop pop 0 setgray Solid {pop []} if 0 setdash} ifelse } def
/BL { stroke userlinewidth 2 mul setlinewidth
      Rounded { 1 setlinejoin 1 setlinecap } if } def
/AL { stroke userlinewidth 2 div setlinewidth
      Rounded { 1 setlinejoin 1 setlinecap } if } def
/UL { dup gnulinewidth mul /userlinewidth exch def
      dup 1 lt {pop 1} if 10 mul /udl exch def } def
/PL { stroke userlinewidth setlinewidth
      Rounded { 1 setlinejoin 1 setlinecap } if } def
/LTw { PL [] 1 setgray } def
/LTb { BL [] 0 0 0 DL } def
/LTa { AL [1 udl mul 2 udl mul] 0 setdash 0 0 0 setrgbcolor } def
/LT0 { PL [] 1 0 0 DL } def
/LT1 { PL [4 dl 2 dl] 0 1 0 DL } def
/LT2 { PL [2 dl 3 dl] 0 0 1 DL } def
/LT3 { PL [1 dl 1.5 dl] 1 0 1 DL } def
/LT4 { PL [5 dl 2 dl 1 dl 2 dl] 0 1 1 DL } def
/LT5 { PL [4 dl 3 dl 1 dl 3 dl] 1 1 0 DL } def
/LT6 { PL [2 dl 2 dl 2 dl 4 dl] 0 0 0 DL } def
/LT7 { PL [2 dl 2 dl 2 dl 2 dl 2 dl 4 dl] 1 0.3 0 DL } def
/LT8 { PL [2 dl 2 dl 2 dl 2 dl 2 dl 2 dl 2 dl 4 dl] 0.5 0.5 0.5 DL } def
/Pnt { stroke [] 0 setdash
   gsave 1 setlinecap M 0 0 V stroke grestore } def
/Dia { stroke [] 0 setdash 2 copy vpt add M
  hpt neg vpt neg V hpt vpt neg V
  hpt vpt V hpt neg vpt V closepath stroke
  Pnt } def
/Pls { stroke [] 0 setdash vpt sub M 0 vpt2 V
  currentpoint stroke M
  hpt neg vpt neg R hpt2 0 V stroke
  } def
/Box { stroke [] 0 setdash 2 copy exch hpt sub exch vpt add M
  0 vpt2 neg V hpt2 0 V 0 vpt2 V
  hpt2 neg 0 V closepath stroke
  Pnt } def
/Crs { stroke [] 0 setdash exch hpt sub exch vpt add M
  hpt2 vpt2 neg V currentpoint stroke M
  hpt2 neg 0 R hpt2 vpt2 V stroke } def
/TriU { stroke [] 0 setdash 2 copy vpt 1.12 mul add M
  hpt neg vpt -1.62 mul V
  hpt 2 mul 0 V
  hpt neg vpt 1.62 mul V closepath stroke
  Pnt  } def
/Star { 2 copy Pls Crs } def
/BoxF { stroke [] 0 setdash exch hpt sub exch vpt add M
  0 vpt2 neg V  hpt2 0 V  0 vpt2 V
  hpt2 neg 0 V  closepath fill } def
/TriUF { stroke [] 0 setdash vpt 1.12 mul add M
  hpt neg vpt -1.62 mul V
  hpt 2 mul 0 V
  hpt neg vpt 1.62 mul V closepath fill } def
/TriD { stroke [] 0 setdash 2 copy vpt 1.12 mul sub M
  hpt neg vpt 1.62 mul V
  hpt 2 mul 0 V
  hpt neg vpt -1.62 mul V closepath stroke
  Pnt  } def
/TriDF { stroke [] 0 setdash vpt 1.12 mul sub M
  hpt neg vpt 1.62 mul V
  hpt 2 mul 0 V
  hpt neg vpt -1.62 mul V closepath fill} def
/DiaF { stroke [] 0 setdash vpt add M
  hpt neg vpt neg V hpt vpt neg V
  hpt vpt V hpt neg vpt V closepath fill } def
/Pent { stroke [] 0 setdash 2 copy gsave
  translate 0 hpt M 4 {72 rotate 0 hpt L} repeat
  closepath stroke grestore Pnt } def
/PentF { stroke [] 0 setdash gsave
  translate 0 hpt M 4 {72 rotate 0 hpt L} repeat
  closepath fill grestore } def
/Circle { stroke [] 0 setdash 2 copy
  hpt 0 360 arc stroke Pnt } def
/CircleF { stroke [] 0 setdash hpt 0 360 arc fill } def
/C0 { BL [] 0 setdash 2 copy moveto vpt 90 450  arc } bind def
/C1 { BL [] 0 setdash 2 copy        moveto
       2 copy  vpt 0 90 arc closepath fill
               vpt 0 360 arc closepath } bind def
/C2 { BL [] 0 setdash 2 copy moveto
       2 copy  vpt 90 180 arc closepath fill
               vpt 0 360 arc closepath } bind def
/C3 { BL [] 0 setdash 2 copy moveto
       2 copy  vpt 0 180 arc closepath fill
               vpt 0 360 arc closepath } bind def
/C4 { BL [] 0 setdash 2 copy moveto
       2 copy  vpt 180 270 arc closepath fill
               vpt 0 360 arc closepath } bind def
/C5 { BL [] 0 setdash 2 copy moveto
       2 copy  vpt 0 90 arc
       2 copy moveto
       2 copy  vpt 180 270 arc closepath fill
               vpt 0 360 arc } bind def
/C6 { BL [] 0 setdash 2 copy moveto
      2 copy  vpt 90 270 arc closepath fill
              vpt 0 360 arc closepath } bind def
/C7 { BL [] 0 setdash 2 copy moveto
      2 copy  vpt 0 270 arc closepath fill
              vpt 0 360 arc closepath } bind def
/C8 { BL [] 0 setdash 2 copy moveto
      2 copy vpt 270 360 arc closepath fill
              vpt 0 360 arc closepath } bind def
/C9 { BL [] 0 setdash 2 copy moveto
      2 copy  vpt 270 450 arc closepath fill
              vpt 0 360 arc closepath } bind def
/C10 { BL [] 0 setdash 2 copy 2 copy moveto vpt 270 360 arc closepath fill
       2 copy moveto
       2 copy vpt 90 180 arc closepath fill
               vpt 0 360 arc closepath } bind def
/C11 { BL [] 0 setdash 2 copy moveto
       2 copy  vpt 0 180 arc closepath fill
       2 copy moveto
       2 copy  vpt 270 360 arc closepath fill
               vpt 0 360 arc closepath } bind def
/C12 { BL [] 0 setdash 2 copy moveto
       2 copy  vpt 180 360 arc closepath fill
               vpt 0 360 arc closepath } bind def
/C13 { BL [] 0 setdash  2 copy moveto
       2 copy  vpt 0 90 arc closepath fill
       2 copy moveto
       2 copy  vpt 180 360 arc closepath fill
               vpt 0 360 arc closepath } bind def
/C14 { BL [] 0 setdash 2 copy moveto
       2 copy  vpt 90 360 arc closepath fill
               vpt 0 360 arc } bind def
/C15 { BL [] 0 setdash 2 copy vpt 0 360 arc closepath fill
               vpt 0 360 arc closepath } bind def
/Rec   { newpath 4 2 roll moveto 1 index 0 rlineto 0 exch rlineto
       neg 0 rlineto closepath } bind def
/Square { dup Rec } bind def
/Bsquare { vpt sub exch vpt sub exch vpt2 Square } bind def
/S0 { BL [] 0 setdash 2 copy moveto 0 vpt rlineto BL Bsquare } bind def
/S1 { BL [] 0 setdash 2 copy vpt Square fill Bsquare } bind def
/S2 { BL [] 0 setdash 2 copy exch vpt sub exch vpt Square fill Bsquare } bind def
/S3 { BL [] 0 setdash 2 copy exch vpt sub exch vpt2 vpt Rec fill Bsquare } bind def
/S4 { BL [] 0 setdash 2 copy exch vpt sub exch vpt sub vpt Square fill Bsquare } bind def
/S5 { BL [] 0 setdash 2 copy 2 copy vpt Square fill
       exch vpt sub exch vpt sub vpt Square fill Bsquare } bind def
/S6 { BL [] 0 setdash 2 copy exch vpt sub exch vpt sub vpt vpt2 Rec fill Bsquare } bind def
/S7 { BL [] 0 setdash 2 copy exch vpt sub exch vpt sub vpt vpt2 Rec fill
       2 copy vpt Square fill
       Bsquare } bind def
/S8 { BL [] 0 setdash 2 copy vpt sub vpt Square fill Bsquare } bind def
/S9 { BL [] 0 setdash 2 copy vpt sub vpt vpt2 Rec fill Bsquare } bind def
/S10 { BL [] 0 setdash 2 copy vpt sub vpt Square fill 2 copy exch vpt sub exch vpt Square fill
       Bsquare } bind def
/S11 { BL [] 0 setdash 2 copy vpt sub vpt Square fill 2 copy exch vpt sub exch vpt2 vpt Rec fill
       Bsquare } bind def
/S12 { BL [] 0 setdash 2 copy exch vpt sub exch vpt sub vpt2 vpt Rec fill Bsquare } bind def
/S13 { BL [] 0 setdash 2 copy exch vpt sub exch vpt sub vpt2 vpt Rec fill
       2 copy vpt Square fill Bsquare } bind def
/S14 { BL [] 0 setdash 2 copy exch vpt sub exch vpt sub vpt2 vpt Rec fill
       2 copy exch vpt sub exch vpt Square fill Bsquare } bind def
/S15 { BL [] 0 setdash 2 copy Bsquare fill Bsquare } bind def
/D0 { gsave translate 45 rotate 0 0 S0 stroke grestore } bind def
/D1 { gsave translate 45 rotate 0 0 S1 stroke grestore } bind def
/D2 { gsave translate 45 rotate 0 0 S2 stroke grestore } bind def
/D3 { gsave translate 45 rotate 0 0 S3 stroke grestore } bind def
/D4 { gsave translate 45 rotate 0 0 S4 stroke grestore } bind def
/D5 { gsave translate 45 rotate 0 0 S5 stroke grestore } bind def
/D6 { gsave translate 45 rotate 0 0 S6 stroke grestore } bind def
/D7 { gsave translate 45 rotate 0 0 S7 stroke grestore } bind def
/D8 { gsave translate 45 rotate 0 0 S8 stroke grestore } bind def
/D9 { gsave translate 45 rotate 0 0 S9 stroke grestore } bind def
/D10 { gsave translate 45 rotate 0 0 S10 stroke grestore } bind def
/D11 { gsave translate 45 rotate 0 0 S11 stroke grestore } bind def
/D12 { gsave translate 45 rotate 0 0 S12 stroke grestore } bind def
/D13 { gsave translate 45 rotate 0 0 S13 stroke grestore } bind def
/D14 { gsave translate 45 rotate 0 0 S14 stroke grestore } bind def
/D15 { gsave translate 45 rotate 0 0 S15 stroke grestore } bind def
/DiaE { stroke [] 0 setdash vpt add M
  hpt neg vpt neg V hpt vpt neg V
  hpt vpt V hpt neg vpt V closepath stroke } def
/BoxE { stroke [] 0 setdash exch hpt sub exch vpt add M
  0 vpt2 neg V hpt2 0 V 0 vpt2 V
  hpt2 neg 0 V closepath stroke } def
/TriUE { stroke [] 0 setdash vpt 1.12 mul add M
  hpt neg vpt -1.62 mul V
  hpt 2 mul 0 V
  hpt neg vpt 1.62 mul V closepath stroke } def
/TriDE { stroke [] 0 setdash vpt 1.12 mul sub M
  hpt neg vpt 1.62 mul V
  hpt 2 mul 0 V
  hpt neg vpt -1.62 mul V closepath stroke } def
/PentE { stroke [] 0 setdash gsave
  translate 0 hpt M 4 {72 rotate 0 hpt L} repeat
  closepath stroke grestore } def
/CircE { stroke [] 0 setdash 
  hpt 0 360 arc stroke } def
/Opaque { gsave closepath 1 setgray fill grestore 0 setgray closepath } def
/DiaW { stroke [] 0 setdash vpt add M
  hpt neg vpt neg V hpt vpt neg V
  hpt vpt V hpt neg vpt V Opaque stroke } def
/BoxW { stroke [] 0 setdash exch hpt sub exch vpt add M
  0 vpt2 neg V hpt2 0 V 0 vpt2 V
  hpt2 neg 0 V Opaque stroke } def
/TriUW { stroke [] 0 setdash vpt 1.12 mul add M
  hpt neg vpt -1.62 mul V
  hpt 2 mul 0 V
  hpt neg vpt 1.62 mul V Opaque stroke } def
/TriDW { stroke [] 0 setdash vpt 1.12 mul sub M
  hpt neg vpt 1.62 mul V
  hpt 2 mul 0 V
  hpt neg vpt -1.62 mul V Opaque stroke } def
/PentW { stroke [] 0 setdash gsave
  translate 0 hpt M 4 {72 rotate 0 hpt L} repeat
  Opaque stroke grestore } def
/CircW { stroke [] 0 setdash 
  hpt 0 360 arc Opaque stroke } def
/BoxFill { gsave Rec 1 setgray fill grestore } def
/BoxColFill {
  gsave Rec
  /Fillden exch def
  currentrgbcolor
  /ColB exch def /ColG exch def /ColR exch def
  /ColR ColR Fillden mul Fillden sub 1 add def
  /ColG ColG Fillden mul Fillden sub 1 add def
  /ColB ColB Fillden mul Fillden sub 1 add def
  ColR ColG ColB setrgbcolor
  fill grestore } def
%
%
/PatternFill { gsave /PFa [ 9 2 roll ] def
    PFa 0 get PFa 2 get 2 div add PFa 1 get PFa 3 get 2 div add translate
    PFa 2 get -2 div PFa 3 get -2 div PFa 2 get PFa 3 get Rec
    gsave 1 setgray fill grestore clip
    currentlinewidth 0.5 mul setlinewidth
    /PFs PFa 2 get dup mul PFa 3 get dup mul add sqrt def
    0 0 M PFa 5 get rotate PFs -2 div dup translate
	0 1 PFs PFa 4 get div 1 add floor cvi
	{ PFa 4 get mul 0 M 0 PFs V } for
    0 PFa 6 get ne {
	0 1 PFs PFa 4 get div 1 add floor cvi
	{ PFa 4 get mul 0 2 1 roll M PFs 0 V } for
    } if
    stroke grestore } def
/Symbol-Oblique /Symbol findfont [1 0 .167 1 0 0] makefont
dup length dict begin {1 index /FID eq {pop pop} {def} ifelse} forall
currentdict end definefont pop
end
gnudict begin
gsave
0 0 translate
0.100 0.100 scale
0 setgray
newpath
1.000 UL
LTb
500 300 M
63 0 V
2887 0 R
-63 0 V
1.000 UL
LTb
500 593 M
63 0 V
2887 0 R
-63 0 V
1.000 UL
LTb
500 887 M
63 0 V
2887 0 R
-63 0 V
1.000 UL
LTb
500 1180 M
63 0 V
2887 0 R
-63 0 V
1.000 UL
LTb
500 1473 M
63 0 V
2887 0 R
-63 0 V
1.000 UL
LTb
500 1767 M
63 0 V
2887 0 R
-63 0 V
1.000 UL
LTb
500 2060 M
63 0 V
2887 0 R
-63 0 V
1.000 UL
LTb
500 300 M
0 63 V
0 1697 R
0 -63 V
1.000 UL
LTb
869 300 M
0 63 V
0 1697 R
0 -63 V
1.000 UL
LTb
1238 300 M
0 63 V
0 1697 R
0 -63 V
1.000 UL
LTb
1606 300 M
0 63 V
0 1697 R
0 -63 V
1.000 UL
LTb
1975 300 M
0 63 V
0 1697 R
0 -63 V
1.000 UL
LTb
2344 300 M
0 63 V
0 1697 R
0 -63 V
1.000 UL
LTb
2713 300 M
0 63 V
0 1697 R
0 -63 V
1.000 UL
LTb
3081 300 M
0 63 V
0 1697 R
0 -63 V
1.000 UL
LTb
3450 300 M
0 63 V
0 1697 R
0 -63 V
1.000 UL
LTb
1.000 UL
LTb
500 300 M
2950 0 V
0 1760 V
-2950 0 V
500 300 L
LTb
LTb
1.000 UP
1.000 UL
LT0
500 1180 M
530 853 L
560 687 L
589 581 L
30 -70 V
30 -45 V
30 -26 V
30 -11 V
29 2 V
30 11 V
30 19 V
30 27 V
30 31 V
29 37 V
30 42 V
30 44 V
30 48 V
30 50 V
29 53 V
30 54 V
30 56 V
30 57 V
30 58 V
29 59 V
30 59 V
30 59 V
30 59 V
30 59 V
29 58 V
30 58 V
30 56 V
30 55 V
30 54 V
29 51 V
30 49 V
30 47 V
30 43 V
30 39 V
29 34 V
30 29 V
30 23 V
30 16 V
30 6 V
29 -4 V
30 -18 V
30 -34 V
30 -57 V
30 -87 V
29 -131 V
30 -215 V
30 -410 V
30 -215 V
29 -131 V
30 -87 V
30 -57 V
30 -34 V
30 -18 V
29 -4 V
30 6 V
30 16 V
30 23 V
30 29 V
29 34 V
30 39 V
30 43 V
30 47 V
30 49 V
29 51 V
30 54 V
30 55 V
30 56 V
30 58 V
29 58 V
30 59 V
30 59 V
30 59 V
30 59 V
29 59 V
30 58 V
30 57 V
30 56 V
30 54 V
29 53 V
30 50 V
30 48 V
30 44 V
30 42 V
29 37 V
30 31 V
30 27 V
30 19 V
30 11 V
29 2 V
30 -11 V
30 -26 V
30 -45 V
30 -70 V
29 -106 V
30 -166 V
30 -327 V
1.000 UL
LT1
500 1180 M
30 -98 V
30 -92 V
29 -83 V
30 -72 V
30 -58 V
30 -45 V
30 -32 V
29 -21 V
30 -10 V
30 -1 V
30 7 V
30 13 V
29 20 V
30 25 V
30 30 V
30 33 V
30 37 V
29 39 V
30 42 V
30 44 V
30 45 V
30 46 V
29 47 V
30 48 V
30 48 V
30 48 V
30 47 V
29 47 V
30 46 V
30 45 V
30 42 V
30 41 V
29 38 V
30 35 V
30 31 V
30 28 V
30 22 V
29 17 V
30 11 V
30 3 V
30 -6 V
30 -15 V
29 -27 V
30 -38 V
30 -52 V
30 -64 V
30 -78 V
29 -88 V
30 -96 V
30 -98 V
30 -96 V
29 -88 V
30 -78 V
30 -64 V
30 -52 V
30 -38 V
29 -27 V
30 -15 V
30 -6 V
30 3 V
30 11 V
29 17 V
30 22 V
30 28 V
30 31 V
30 35 V
29 38 V
30 41 V
30 42 V
30 45 V
30 46 V
29 47 V
30 47 V
30 48 V
30 48 V
30 48 V
29 47 V
30 46 V
30 45 V
30 44 V
30 42 V
29 39 V
30 37 V
30 33 V
30 30 V
30 25 V
29 20 V
30 13 V
30 7 V
30 -1 V
30 -10 V
29 -21 V
30 -32 V
30 -45 V
30 -58 V
30 -72 V
29 -83 V
30 -92 V
30 -98 V
1.000 UL
LT2
500 1180 M
30 -60 V
30 -58 V
29 -55 V
30 -49 V
30 -42 V
30 -36 V
30 -28 V
29 -20 V
30 -14 V
30 -7 V
30 0 V
30 5 V
29 11 V
30 16 V
30 19 V
30 23 V
30 27 V
29 29 V
30 31 V
30 34 V
30 35 V
30 36 V
29 37 V
30 38 V
30 37 V
30 38 V
30 38 V
29 36 V
30 36 V
30 34 V
30 33 V
30 30 V
29 28 V
30 25 V
30 21 V
30 18 V
30 13 V
29 8 V
30 3 V
30 -4 V
30 -10 V
30 -17 V
29 -24 V
30 -32 V
30 -39 V
30 -46 V
30 -52 V
29 -56 V
30 -60 V
30 -60 V
30 -60 V
29 -56 V
30 -52 V
30 -46 V
30 -39 V
30 -32 V
29 -24 V
30 -17 V
30 -10 V
30 -4 V
30 3 V
29 8 V
30 13 V
30 18 V
30 21 V
30 25 V
29 28 V
30 30 V
30 33 V
30 34 V
30 36 V
29 36 V
30 38 V
30 38 V
30 37 V
30 38 V
29 37 V
30 36 V
30 35 V
30 34 V
30 31 V
29 29 V
30 27 V
30 23 V
30 19 V
30 16 V
29 11 V
30 5 V
30 0 V
30 -7 V
30 -14 V
29 -20 V
30 -28 V
30 -36 V
30 -42 V
30 -49 V
29 -55 V
30 -58 V
30 -60 V
1.000 UL
LT3
500 1180 M
30 -40 V
30 -38 V
29 -37 V
30 -34 V
30 -30 V
30 -26 V
30 -22 V
29 -16 V
30 -13 V
30 -7 V
30 -3 V
30 2 V
29 6 V
30 9 V
30 13 V
30 16 V
30 18 V
29 21 V
30 24 V
30 24 V
30 27 V
30 27 V
29 28 V
30 29 V
30 29 V
30 29 V
30 29 V
29 28 V
30 27 V
30 25 V
30 24 V
30 22 V
29 20 V
30 18 V
30 14 V
30 11 V
30 8 V
29 3 V
30 0 V
30 -5 V
30 -10 V
30 -14 V
29 -20 V
30 -23 V
30 -29 V
30 -32 V
30 -35 V
29 -38 V
30 -39 V
30 -40 V
30 -39 V
29 -38 V
30 -35 V
30 -32 V
30 -29 V
30 -23 V
29 -20 V
30 -14 V
30 -10 V
30 -5 V
30 0 V
29 3 V
30 8 V
30 11 V
30 14 V
30 18 V
29 20 V
30 22 V
30 24 V
30 25 V
30 27 V
29 28 V
30 29 V
30 29 V
30 29 V
30 29 V
29 28 V
30 27 V
30 27 V
30 24 V
30 24 V
29 21 V
30 18 V
30 16 V
30 13 V
30 9 V
29 6 V
30 2 V
30 -3 V
30 -7 V
30 -13 V
29 -16 V
30 -22 V
30 -26 V
30 -30 V
30 -34 V
29 -37 V
30 -38 V
30 -40 V
1.000 UL
LT4
500 1180 M
30 -26 V
30 -26 V
29 -25 V
30 -23 V
30 -21 V
30 -18 V
30 -15 V
29 -13 V
30 -9 V
30 -7 V
30 -3 V
30 0 V
29 3 V
30 5 V
30 9 V
30 10 V
30 13 V
29 15 V
30 16 V
30 18 V
30 19 V
30 20 V
29 21 V
30 21 V
30 21 V
30 22 V
30 21 V
29 20 V
30 19 V
30 19 V
30 17 V
30 16 V
29 14 V
30 11 V
30 10 V
30 7 V
30 4 V
29 1 V
30 -1 V
30 -5 V
30 -8 V
30 -11 V
29 -14 V
30 -17 V
30 -20 V
30 -22 V
30 -23 V
29 -26 V
30 -26 V
30 -26 V
30 -26 V
29 -26 V
30 -23 V
30 -22 V
30 -20 V
30 -17 V
29 -14 V
30 -11 V
30 -8 V
30 -5 V
30 -1 V
29 1 V
30 4 V
30 7 V
30 10 V
30 11 V
29 14 V
30 16 V
30 17 V
30 19 V
30 19 V
29 20 V
30 21 V
30 22 V
30 21 V
30 21 V
29 21 V
30 20 V
30 19 V
30 18 V
30 16 V
29 15 V
30 13 V
30 10 V
30 9 V
30 5 V
29 3 V
30 0 V
30 -3 V
30 -7 V
30 -9 V
29 -13 V
30 -15 V
30 -18 V
30 -21 V
30 -23 V
29 -25 V
30 -26 V
30 -26 V
1.000 UL
LTb
500 300 M
2950 0 V
0 1760 V
-2950 0 V
500 300 L
1.000 UP
1.000 UL
LTb
1372 524 M
63 0 V
647 0 R
-63 0 V
1.000 UL
LTb
1372 638 M
63 0 V
647 0 R
-63 0 V
1.000 UL
LTb
1372 752 M
63 0 V
647 0 R
-63 0 V
1.000 UL
LTb
1372 866 M
63 0 V
647 0 R
-63 0 V
1.000 UL
LTb
1372 980 M
63 0 V
647 0 R
-63 0 V
1.000 UL
LTb
1372 524 M
0 63 V
0 393 R
0 -63 V
1.000 UL
LTb
1550 524 M
0 63 V
0 393 R
0 -63 V
1.000 UL
LTb
1727 524 M
0 63 V
0 393 R
0 -63 V
1.000 UL
LTb
1905 524 M
0 63 V
0 393 R
0 -63 V
1.000 UL
LTb
2082 524 M
0 63 V
0 393 R
0 -63 V
1.000 UL
LTb
1.000 UL
LTb
1372 524 M
710 0 V
0 456 V
-710 0 V
0 -456 V
1.000 UP
1.000 UL
LT0
1405 980 M
3 -1 V
7 -2 V
7 -3 V
7 -2 V
8 -3 V
7 -2 V
7 -3 V
7 -2 V
7 -3 V
7 -3 V
8 -3 V
7 -3 V
7 -3 V
7 -3 V
7 -4 V
7 -3 V
8 -3 V
7 -4 V
7 -4 V
7 -3 V
7 -4 V
7 -4 V
8 -4 V
7 -4 V
7 -5 V
7 -4 V
7 -5 V
7 -5 V
8 -5 V
7 -5 V
7 -5 V
7 -6 V
7 -5 V
8 -6 V
7 -7 V
7 -6 V
7 -7 V
7 -7 V
7 -8 V
8 -8 V
7 -8 V
7 -9 V
7 -10 V
7 -12 V
7 -13 V
8 -18 V
7 -13 V
7 -12 V
7 -10 V
7 -9 V
7 -8 V
8 -8 V
7 -8 V
7 -7 V
7 -7 V
7 -6 V
7 -7 V
8 -6 V
7 -5 V
7 -6 V
7 -5 V
7 -5 V
8 -5 V
7 -5 V
7 -5 V
7 -4 V
7 -5 V
7 -4 V
8 -4 V
7 -4 V
7 -4 V
7 -3 V
7 -4 V
7 -4 V
8 -3 V
7 -3 V
7 -4 V
7 -3 V
7 -3 V
7 -3 V
8 -3 V
7 -3 V
7 -3 V
7 -2 V
7 -3 V
7 -2 V
8 -3 V
7 -2 V
7 -3 V
7 -2 V
3 -1 V
1.000 UL
LT1
1372 863 M
7 -2 V
7 -2 V
8 -2 V
7 -2 V
7 -2 V
7 -2 V
7 -2 V
7 -2 V
8 -2 V
7 -2 V
7 -2 V
7 -2 V
7 -2 V
7 -2 V
8 -2 V
7 -3 V
7 -2 V
7 -2 V
7 -2 V
7 -2 V
8 -3 V
7 -2 V
7 -2 V
7 -2 V
7 -3 V
7 -2 V
8 -2 V
7 -3 V
7 -2 V
7 -2 V
7 -3 V
7 -2 V
8 -2 V
7 -3 V
7 -2 V
7 -2 V
7 -3 V
8 -2 V
7 -3 V
7 -2 V
7 -2 V
7 -3 V
7 -2 V
8 -3 V
7 -2 V
7 -2 V
7 -3 V
7 -2 V
7 -3 V
8 -2 V
7 -3 V
7 -2 V
7 -3 V
7 -2 V
7 -2 V
8 -3 V
7 -2 V
7 -3 V
7 -2 V
7 -2 V
7 -3 V
8 -2 V
7 -3 V
7 -2 V
7 -2 V
7 -3 V
8 -2 V
7 -2 V
7 -3 V
7 -2 V
7 -2 V
7 -3 V
8 -2 V
7 -2 V
7 -3 V
7 -2 V
7 -2 V
7 -2 V
8 -3 V
7 -2 V
7 -2 V
7 -2 V
7 -2 V
7 -3 V
8 -2 V
7 -2 V
7 -2 V
7 -2 V
7 -2 V
7 -2 V
8 -2 V
7 -2 V
7 -2 V
7 -2 V
7 -2 V
7 -2 V
8 -2 V
7 -2 V
7 -2 V
1.000 UL
LT2
1372 822 M
7 -1 V
7 -1 V
8 -2 V
7 -1 V
7 -1 V
7 -2 V
7 -1 V
7 -1 V
8 -2 V
7 -1 V
7 -1 V
7 -2 V
7 -1 V
7 -1 V
8 -2 V
7 -1 V
7 -1 V
7 -2 V
7 -1 V
7 -2 V
8 -1 V
7 -2 V
7 -1 V
7 -1 V
7 -2 V
7 -1 V
8 -2 V
7 -1 V
7 -2 V
7 -1 V
7 -2 V
7 -1 V
8 -1 V
7 -2 V
7 -1 V
7 -2 V
7 -1 V
8 -2 V
7 -1 V
7 -2 V
7 -1 V
7 -2 V
7 -1 V
8 -2 V
7 -1 V
7 -2 V
7 -1 V
7 -2 V
7 -1 V
8 -2 V
7 -1 V
7 -2 V
7 -1 V
7 -2 V
7 -1 V
8 -2 V
7 -1 V
7 -2 V
7 -1 V
7 -2 V
7 -1 V
8 -2 V
7 -1 V
7 -2 V
7 -1 V
7 -2 V
8 -1 V
7 -1 V
7 -2 V
7 -1 V
7 -2 V
7 -1 V
8 -2 V
7 -1 V
7 -2 V
7 -1 V
7 -1 V
7 -2 V
8 -1 V
7 -2 V
7 -1 V
7 -2 V
7 -1 V
7 -1 V
8 -2 V
7 -1 V
7 -1 V
7 -2 V
7 -1 V
7 -1 V
8 -2 V
7 -1 V
7 -1 V
7 -2 V
7 -1 V
7 -1 V
8 -2 V
7 -1 V
7 -1 V
1.000 UL
LT3
1372 799 M
7 -1 V
7 -1 V
8 -1 V
7 -1 V
7 -1 V
7 0 V
7 -1 V
7 -1 V
8 -1 V
7 -1 V
7 -1 V
7 -1 V
7 -1 V
7 -1 V
8 -1 V
7 -1 V
7 -1 V
7 0 V
7 -1 V
7 -1 V
8 -1 V
7 -1 V
7 -1 V
7 -1 V
7 -1 V
7 -1 V
8 -1 V
7 -1 V
7 -1 V
7 -1 V
7 -1 V
7 -1 V
8 -1 V
7 -1 V
7 -1 V
7 -1 V
7 -1 V
8 -1 V
7 -1 V
7 -1 V
7 -1 V
7 -1 V
7 -1 V
8 -1 V
7 -1 V
7 -1 V
7 -1 V
7 -1 V
7 -1 V
8 0 V
7 -1 V
7 -1 V
7 -1 V
7 -1 V
7 -1 V
8 -1 V
7 -1 V
7 -1 V
7 -1 V
7 -1 V
7 -1 V
8 -1 V
7 -1 V
7 -1 V
7 -1 V
7 -1 V
8 -1 V
7 -1 V
7 -1 V
7 -1 V
7 -1 V
7 -1 V
8 -1 V
7 -1 V
7 -1 V
7 -1 V
7 -1 V
7 -1 V
8 -1 V
7 -1 V
7 -1 V
7 0 V
7 -1 V
7 -1 V
8 -1 V
7 -1 V
7 -1 V
7 -1 V
7 -1 V
7 -1 V
8 -1 V
7 -1 V
7 -1 V
7 0 V
7 -1 V
7 -1 V
8 -1 V
7 -1 V
7 -1 V
1.000 UL
LT4
1372 783 M
7 0 V
7 -1 V
8 -1 V
7 0 V
7 -1 V
7 0 V
7 -1 V
7 -1 V
8 0 V
7 -1 V
7 0 V
7 -1 V
7 -1 V
7 0 V
8 -1 V
7 0 V
7 -1 V
7 -1 V
7 0 V
7 -1 V
8 -1 V
7 0 V
7 -1 V
7 0 V
7 -1 V
7 -1 V
8 0 V
7 -1 V
7 -1 V
7 0 V
7 -1 V
7 -1 V
8 0 V
7 -1 V
7 -1 V
7 0 V
7 -1 V
8 0 V
7 -1 V
7 -1 V
7 0 V
7 -1 V
7 -1 V
8 0 V
7 -1 V
7 -1 V
7 0 V
7 -1 V
7 -1 V
8 0 V
7 -1 V
7 -1 V
7 0 V
7 -1 V
7 -1 V
8 0 V
7 -1 V
7 -1 V
7 0 V
7 -1 V
7 -1 V
8 0 V
7 -1 V
7 0 V
7 -1 V
7 -1 V
8 0 V
7 -1 V
7 -1 V
7 0 V
7 -1 V
7 -1 V
8 0 V
7 -1 V
7 -1 V
7 0 V
7 -1 V
7 0 V
8 -1 V
7 -1 V
7 0 V
7 -1 V
7 -1 V
7 0 V
8 -1 V
7 0 V
7 -1 V
7 -1 V
7 0 V
7 -1 V
8 0 V
7 -1 V
7 -1 V
7 0 V
7 -1 V
7 0 V
8 -1 V
7 -1 V
7 0 V
1.000 UL
LTb
1372 524 M
710 0 V
0 456 V
-710 0 V
0 -456 V
1.000 UP
stroke
grestore
end
showpage
}}%
\put(2082,424){\makebox(0,0){ 0.2}}%
\put(1905,424){\makebox(0,0){ 0.1}}%
\put(1727,424){\makebox(0,0){ 0}}%
\put(1550,424){\makebox(0,0){-0.1}}%
\put(1372,424){\makebox(0,0){-0.2}}%
\put(1322,980){\makebox(0,0)[r]{ 0.1}}%
\put(1322,866){\makebox(0,0)[r]{ 0.05}}%
\put(1322,752){\makebox(0,0)[r]{ 0}}%
\put(1322,638){\makebox(0,0)[r]{-0.05}}%
\put(1322,524){\makebox(0,0)[r]{-0.1}}%
\put(1975,50){\makebox(0,0){angle $\alpha$}}%
\put(100,1180){%
\special{ps: gsave currentpoint currentpoint translate
270 rotate neg exch neg exch translate}%
\makebox(0,0)[b]{\shortstack{torque $(L/\hbar c)\tau_z$}}%
\special{ps: currentpoint grestore moveto}%
}%
\put(3450,200){\makebox(0,0){$\pi$}}%
\put(3081,200){\makebox(0,0){$3\pi/4$}}%
\put(2713,200){\makebox(0,0){$\pi/2$}}%
\put(2344,200){\makebox(0,0){$\pi/4$}}%
\put(1975,200){\makebox(0,0){0}}%
\put(1606,200){\makebox(0,0){$-\pi/4$}}%
\put(1238,200){\makebox(0,0){$-\pi/2$}}%
\put(869,200){\makebox(0,0){$-3\pi/4$}}%
\put(500,200){\makebox(0,0){$-\pi$}}%
\put(450,2060){\makebox(0,0)[r]{ 0.15}}%
\put(450,1767){\makebox(0,0)[r]{ 0.1}}%
\put(450,1473){\makebox(0,0)[r]{ 0.05}}%
\put(450,1180){\makebox(0,0)[r]{ 0}}%
\put(450,887){\makebox(0,0)[r]{-0.05}}%
\put(450,593){\makebox(0,0)[r]{-0.1}}%
\put(450,300){\makebox(0,0)[r]{-0.15}}%
\end{picture}%
\endgroup
 

%% file: dicroic.tex
\begingroup%
  \makeatletter%
  \newcommand{\GNUPLOTspecial}{%
    \@sanitize\catcode`\%=14\relax\special}%
  \setlength{\unitlength}{0.1bp}%
\begin{picture}(3600,2160)(0,0)%
{\GNUPLOTspecial{"
/gnudict 256 dict def
gnudict begin
/Color false def
/Solid false def
/gnulinewidth 5.000 def
/userlinewidth gnulinewidth def
/vshift -33 def
/dl {10.0 mul} def
/hpt_ 31.5 def
/vpt_ 31.5 def
/hpt hpt_ def
/vpt vpt_ def
/Rounded false def
/M {moveto} bind def
/L {lineto} bind def
/R {rmoveto} bind def
/V {rlineto} bind def
/N {newpath moveto} bind def
/C {setrgbcolor} bind def
/f {rlineto fill} bind def
/vpt2 vpt 2 mul def
/hpt2 hpt 2 mul def
/Lshow { currentpoint stroke M
  0 vshift R show } def
/Rshow { currentpoint stroke M
  dup stringwidth pop neg vshift R show } def
/Cshow { currentpoint stroke M
  dup stringwidth pop -2 div vshift R show } def
/UP { dup vpt_ mul /vpt exch def hpt_ mul /hpt exch def
  /hpt2 hpt 2 mul def /vpt2 vpt 2 mul def } def
/DL { Color {setrgbcolor Solid {pop []} if 0 setdash }
 {pop pop pop 0 setgray Solid {pop []} if 0 setdash} ifelse } def
/BL { stroke userlinewidth 2 mul setlinewidth
      Rounded { 1 setlinejoin 1 setlinecap } if } def
/AL { stroke userlinewidth 2 div setlinewidth
      Rounded { 1 setlinejoin 1 setlinecap } if } def
/UL { dup gnulinewidth mul /userlinewidth exch def
      dup 1 lt {pop 1} if 10 mul /udl exch def } def
/PL { stroke userlinewidth setlinewidth
      Rounded { 1 setlinejoin 1 setlinecap } if } def
/LTw { PL [] 1 setgray } def
/LTb { BL [] 0 0 0 DL } def
/LTa { AL [1 udl mul 2 udl mul] 0 setdash 0 0 0 setrgbcolor } def
/LT0 { PL [] 1 0 0 DL } def
/LT1 { PL [4 dl 2 dl] 0 1 0 DL } def
/LT2 { PL [2 dl 3 dl] 0 0 1 DL } def
/LT3 { PL [1 dl 1.5 dl] 1 0 1 DL } def
/LT4 { PL [5 dl 2 dl 1 dl 2 dl] 0 1 1 DL } def
/LT5 { PL [4 dl 3 dl 1 dl 3 dl] 1 1 0 DL } def
/LT6 { PL [2 dl 2 dl 2 dl 4 dl] 0 0 0 DL } def
/LT7 { PL [2 dl 2 dl 2 dl 2 dl 2 dl 4 dl] 1 0.3 0 DL } def
/LT8 { PL [2 dl 2 dl 2 dl 2 dl 2 dl 2 dl 2 dl 4 dl] 0.5 0.5 0.5 DL } def
/Pnt { stroke [] 0 setdash
   gsave 1 setlinecap M 0 0 V stroke grestore } def
/Dia { stroke [] 0 setdash 2 copy vpt add M
  hpt neg vpt neg V hpt vpt neg V
  hpt vpt V hpt neg vpt V closepath stroke
  Pnt } def
/Pls { stroke [] 0 setdash vpt sub M 0 vpt2 V
  currentpoint stroke M
  hpt neg vpt neg R hpt2 0 V stroke
  } def
/Box { stroke [] 0 setdash 2 copy exch hpt sub exch vpt add M
  0 vpt2 neg V hpt2 0 V 0 vpt2 V
  hpt2 neg 0 V closepath stroke
  Pnt } def
/Crs { stroke [] 0 setdash exch hpt sub exch vpt add M
  hpt2 vpt2 neg V currentpoint stroke M
  hpt2 neg 0 R hpt2 vpt2 V stroke } def
/TriU { stroke [] 0 setdash 2 copy vpt 1.12 mul add M
  hpt neg vpt -1.62 mul V
  hpt 2 mul 0 V
  hpt neg vpt 1.62 mul V closepath stroke
  Pnt  } def
/Star { 2 copy Pls Crs } def
/BoxF { stroke [] 0 setdash exch hpt sub exch vpt add M
  0 vpt2 neg V  hpt2 0 V  0 vpt2 V
  hpt2 neg 0 V  closepath fill } def
/TriUF { stroke [] 0 setdash vpt 1.12 mul add M
  hpt neg vpt -1.62 mul V
  hpt 2 mul 0 V
  hpt neg vpt 1.62 mul V closepath fill } def
/TriD { stroke [] 0 setdash 2 copy vpt 1.12 mul sub M
  hpt neg vpt 1.62 mul V
  hpt 2 mul 0 V
  hpt neg vpt -1.62 mul V closepath stroke
  Pnt  } def
/TriDF { stroke [] 0 setdash vpt 1.12 mul sub M
  hpt neg vpt 1.62 mul V
  hpt 2 mul 0 V
  hpt neg vpt -1.62 mul V closepath fill} def
/DiaF { stroke [] 0 setdash vpt add M
  hpt neg vpt neg V hpt vpt neg V
  hpt vpt V hpt neg vpt V closepath fill } def
/Pent { stroke [] 0 setdash 2 copy gsave
  translate 0 hpt M 4 {72 rotate 0 hpt L} repeat
  closepath stroke grestore Pnt } def
/PentF { stroke [] 0 setdash gsave
  translate 0 hpt M 4 {72 rotate 0 hpt L} repeat
  closepath fill grestore } def
/Circle { stroke [] 0 setdash 2 copy
  hpt 0 360 arc stroke Pnt } def
/CircleF { stroke [] 0 setdash hpt 0 360 arc fill } def
/C0 { BL [] 0 setdash 2 copy moveto vpt 90 450  arc } bind def
/C1 { BL [] 0 setdash 2 copy        moveto
       2 copy  vpt 0 90 arc closepath fill
               vpt 0 360 arc closepath } bind def
/C2 { BL [] 0 setdash 2 copy moveto
       2 copy  vpt 90 180 arc closepath fill
               vpt 0 360 arc closepath } bind def
/C3 { BL [] 0 setdash 2 copy moveto
       2 copy  vpt 0 180 arc closepath fill
               vpt 0 360 arc closepath } bind def
/C4 { BL [] 0 setdash 2 copy moveto
       2 copy  vpt 180 270 arc closepath fill
               vpt 0 360 arc closepath } bind def
/C5 { BL [] 0 setdash 2 copy moveto
       2 copy  vpt 0 90 arc
       2 copy moveto
       2 copy  vpt 180 270 arc closepath fill
               vpt 0 360 arc } bind def
/C6 { BL [] 0 setdash 2 copy moveto
      2 copy  vpt 90 270 arc closepath fill
              vpt 0 360 arc closepath } bind def
/C7 { BL [] 0 setdash 2 copy moveto
      2 copy  vpt 0 270 arc closepath fill
              vpt 0 360 arc closepath } bind def
/C8 { BL [] 0 setdash 2 copy moveto
      2 copy vpt 270 360 arc closepath fill
              vpt 0 360 arc closepath } bind def
/C9 { BL [] 0 setdash 2 copy moveto
      2 copy  vpt 270 450 arc closepath fill
              vpt 0 360 arc closepath } bind def
/C10 { BL [] 0 setdash 2 copy 2 copy moveto vpt 270 360 arc closepath fill
       2 copy moveto
       2 copy vpt 90 180 arc closepath fill
               vpt 0 360 arc closepath } bind def
/C11 { BL [] 0 setdash 2 copy moveto
       2 copy  vpt 0 180 arc closepath fill
       2 copy moveto
       2 copy  vpt 270 360 arc closepath fill
               vpt 0 360 arc closepath } bind def
/C12 { BL [] 0 setdash 2 copy moveto
       2 copy  vpt 180 360 arc closepath fill
               vpt 0 360 arc closepath } bind def
/C13 { BL [] 0 setdash  2 copy moveto
       2 copy  vpt 0 90 arc closepath fill
       2 copy moveto
       2 copy  vpt 180 360 arc closepath fill
               vpt 0 360 arc closepath } bind def
/C14 { BL [] 0 setdash 2 copy moveto
       2 copy  vpt 90 360 arc closepath fill
               vpt 0 360 arc } bind def
/C15 { BL [] 0 setdash 2 copy vpt 0 360 arc closepath fill
               vpt 0 360 arc closepath } bind def
/Rec   { newpath 4 2 roll moveto 1 index 0 rlineto 0 exch rlineto
       neg 0 rlineto closepath } bind def
/Square { dup Rec } bind def
/Bsquare { vpt sub exch vpt sub exch vpt2 Square } bind def
/S0 { BL [] 0 setdash 2 copy moveto 0 vpt rlineto BL Bsquare } bind def
/S1 { BL [] 0 setdash 2 copy vpt Square fill Bsquare } bind def
/S2 { BL [] 0 setdash 2 copy exch vpt sub exch vpt Square fill Bsquare } bind def
/S3 { BL [] 0 setdash 2 copy exch vpt sub exch vpt2 vpt Rec fill Bsquare } bind def
/S4 { BL [] 0 setdash 2 copy exch vpt sub exch vpt sub vpt Square fill Bsquare } bind def
/S5 { BL [] 0 setdash 2 copy 2 copy vpt Square fill
       exch vpt sub exch vpt sub vpt Square fill Bsquare } bind def
/S6 { BL [] 0 setdash 2 copy exch vpt sub exch vpt sub vpt vpt2 Rec fill Bsquare } bind def
/S7 { BL [] 0 setdash 2 copy exch vpt sub exch vpt sub vpt vpt2 Rec fill
       2 copy vpt Square fill
       Bsquare } bind def
/S8 { BL [] 0 setdash 2 copy vpt sub vpt Square fill Bsquare } bind def
/S9 { BL [] 0 setdash 2 copy vpt sub vpt vpt2 Rec fill Bsquare } bind def
/S10 { BL [] 0 setdash 2 copy vpt sub vpt Square fill 2 copy exch vpt sub exch vpt Square fill
       Bsquare } bind def
/S11 { BL [] 0 setdash 2 copy vpt sub vpt Square fill 2 copy exch vpt sub exch vpt2 vpt Rec fill
       Bsquare } bind def
/S12 { BL [] 0 setdash 2 copy exch vpt sub exch vpt sub vpt2 vpt Rec fill Bsquare } bind def
/S13 { BL [] 0 setdash 2 copy exch vpt sub exch vpt sub vpt2 vpt Rec fill
       2 copy vpt Square fill Bsquare } bind def
/S14 { BL [] 0 setdash 2 copy exch vpt sub exch vpt sub vpt2 vpt Rec fill
       2 copy exch vpt sub exch vpt Square fill Bsquare } bind def
/S15 { BL [] 0 setdash 2 copy Bsquare fill Bsquare } bind def
/D0 { gsave translate 45 rotate 0 0 S0 stroke grestore } bind def
/D1 { gsave translate 45 rotate 0 0 S1 stroke grestore } bind def
/D2 { gsave translate 45 rotate 0 0 S2 stroke grestore } bind def
/D3 { gsave translate 45 rotate 0 0 S3 stroke grestore } bind def
/D4 { gsave translate 45 rotate 0 0 S4 stroke grestore } bind def
/D5 { gsave translate 45 rotate 0 0 S5 stroke grestore } bind def
/D6 { gsave translate 45 rotate 0 0 S6 stroke grestore } bind def
/D7 { gsave translate 45 rotate 0 0 S7 stroke grestore } bind def
/D8 { gsave translate 45 rotate 0 0 S8 stroke grestore } bind def
/D9 { gsave translate 45 rotate 0 0 S9 stroke grestore } bind def
/D10 { gsave translate 45 rotate 0 0 S10 stroke grestore } bind def
/D11 { gsave translate 45 rotate 0 0 S11 stroke grestore } bind def
/D12 { gsave translate 45 rotate 0 0 S12 stroke grestore } bind def
/D13 { gsave translate 45 rotate 0 0 S13 stroke grestore } bind def
/D14 { gsave translate 45 rotate 0 0 S14 stroke grestore } bind def
/D15 { gsave translate 45 rotate 0 0 S15 stroke grestore } bind def
/DiaE { stroke [] 0 setdash vpt add M
  hpt neg vpt neg V hpt vpt neg V
  hpt vpt V hpt neg vpt V closepath stroke } def
/BoxE { stroke [] 0 setdash exch hpt sub exch vpt add M
  0 vpt2 neg V hpt2 0 V 0 vpt2 V
  hpt2 neg 0 V closepath stroke } def
/TriUE { stroke [] 0 setdash vpt 1.12 mul add M
  hpt neg vpt -1.62 mul V
  hpt 2 mul 0 V
  hpt neg vpt 1.62 mul V closepath stroke } def
/TriDE { stroke [] 0 setdash vpt 1.12 mul sub M
  hpt neg vpt 1.62 mul V
  hpt 2 mul 0 V
  hpt neg vpt -1.62 mul V closepath stroke } def
/PentE { stroke [] 0 setdash gsave
  translate 0 hpt M 4 {72 rotate 0 hpt L} repeat
  closepath stroke grestore } def
/CircE { stroke [] 0 setdash 
  hpt 0 360 arc stroke } def
/Opaque { gsave closepath 1 setgray fill grestore 0 setgray closepath } def
/DiaW { stroke [] 0 setdash vpt add M
  hpt neg vpt neg V hpt vpt neg V
  hpt vpt V hpt neg vpt V Opaque stroke } def
/BoxW { stroke [] 0 setdash exch hpt sub exch vpt add M
  0 vpt2 neg V hpt2 0 V 0 vpt2 V
  hpt2 neg 0 V Opaque stroke } def
/TriUW { stroke [] 0 setdash vpt 1.12 mul add M
  hpt neg vpt -1.62 mul V
  hpt 2 mul 0 V
  hpt neg vpt 1.62 mul V Opaque stroke } def
/TriDW { stroke [] 0 setdash vpt 1.12 mul sub M
  hpt neg vpt 1.62 mul V
  hpt 2 mul 0 V
  hpt neg vpt -1.62 mul V Opaque stroke } def
/PentW { stroke [] 0 setdash gsave
  translate 0 hpt M 4 {72 rotate 0 hpt L} repeat
  Opaque stroke grestore } def
/CircW { stroke [] 0 setdash 
  hpt 0 360 arc Opaque stroke } def
/BoxFill { gsave Rec 1 setgray fill grestore } def
/BoxColFill {
  gsave Rec
  /Fillden exch def
  currentrgbcolor
  /ColB exch def /ColG exch def /ColR exch def
  /ColR ColR Fillden mul Fillden sub 1 add def
  /ColG ColG Fillden mul Fillden sub 1 add def
  /ColB ColB Fillden mul Fillden sub 1 add def
  ColR ColG ColB setrgbcolor
  fill grestore } def
%
%
/PatternFill { gsave /PFa [ 9 2 roll ] def
    PFa 0 get PFa 2 get 2 div add PFa 1 get PFa 3 get 2 div add translate
    PFa 2 get -2 div PFa 3 get -2 div PFa 2 get PFa 3 get Rec
    gsave 1 setgray fill grestore clip
    currentlinewidth 0.5 mul setlinewidth
    /PFs PFa 2 get dup mul PFa 3 get dup mul add sqrt def
    0 0 M PFa 5 get rotate PFs -2 div dup translate
	0 1 PFs PFa 4 get div 1 add floor cvi
	{ PFa 4 get mul 0 M 0 PFs V } for
    0 PFa 6 get ne {
	0 1 PFs PFa 4 get div 1 add floor cvi
	{ PFa 4 get mul 0 2 1 roll M PFs 0 V } for
    } if
    stroke grestore } def
/Symbol-Oblique /Symbol findfont [1 0 .167 1 0 0] makefont
dup length dict begin {1 index /FID eq {pop pop} {def} ifelse} forall
currentdict end definefont pop
end
gnudict begin
gsave
0 0 translate
0.100 0.100 scale
0 setgray
newpath
1.000 UL
LTb
550 300 M
63 0 V
2837 0 R
-63 0 V
1.000 UL
LTb
550 477 M
31 0 V
2869 0 R
-31 0 V
550 580 M
31 0 V
2869 0 R
-31 0 V
550 653 M
31 0 V
2869 0 R
-31 0 V
550 710 M
31 0 V
2869 0 R
-31 0 V
550 757 M
31 0 V
2869 0 R
-31 0 V
550 796 M
31 0 V
2869 0 R
-31 0 V
550 830 M
31 0 V
2869 0 R
-31 0 V
550 860 M
31 0 V
2869 0 R
-31 0 V
550 887 M
63 0 V
2837 0 R
-63 0 V
1.000 UL
LTb
550 1063 M
31 0 V
2869 0 R
-31 0 V
550 1167 M
31 0 V
2869 0 R
-31 0 V
550 1240 M
31 0 V
2869 0 R
-31 0 V
550 1297 M
31 0 V
2869 0 R
-31 0 V
550 1343 M
31 0 V
2869 0 R
-31 0 V
550 1382 M
31 0 V
2869 0 R
-31 0 V
550 1416 M
31 0 V
2869 0 R
-31 0 V
550 1446 M
31 0 V
2869 0 R
-31 0 V
550 1473 M
63 0 V
2837 0 R
-63 0 V
1.000 UL
LTb
550 1650 M
31 0 V
2869 0 R
-31 0 V
550 1753 M
31 0 V
2869 0 R
-31 0 V
550 1827 M
31 0 V
2869 0 R
-31 0 V
550 1883 M
31 0 V
2869 0 R
-31 0 V
550 1930 M
31 0 V
2869 0 R
-31 0 V
550 1969 M
31 0 V
2869 0 R
-31 0 V
550 2003 M
31 0 V
2869 0 R
-31 0 V
550 2033 M
31 0 V
2869 0 R
-31 0 V
550 2060 M
63 0 V
2837 0 R
-63 0 V
1.000 UL
LTb
550 300 M
0 63 V
0 1697 R
0 -63 V
1.000 UL
LTb
768 300 M
0 31 V
0 1729 R
0 -31 V
896 300 M
0 31 V
0 1729 R
0 -31 V
986 300 M
0 31 V
0 1729 R
0 -31 V
1057 300 M
0 31 V
0 1729 R
0 -31 V
1114 300 M
0 31 V
0 1729 R
0 -31 V
1163 300 M
0 31 V
0 1729 R
0 -31 V
1205 300 M
0 31 V
0 1729 R
0 -31 V
1242 300 M
0 31 V
0 1729 R
0 -31 V
1275 300 M
0 63 V
0 1697 R
0 -63 V
1.000 UL
LTb
1493 300 M
0 31 V
0 1729 R
0 -31 V
1621 300 M
0 31 V
0 1729 R
0 -31 V
1711 300 M
0 31 V
0 1729 R
0 -31 V
1782 300 M
0 31 V
0 1729 R
0 -31 V
1839 300 M
0 31 V
0 1729 R
0 -31 V
1888 300 M
0 31 V
0 1729 R
0 -31 V
1930 300 M
0 31 V
0 1729 R
0 -31 V
1967 300 M
0 31 V
0 1729 R
0 -31 V
2000 300 M
0 63 V
0 1697 R
0 -63 V
1.000 UL
LTb
2218 300 M
0 31 V
0 1729 R
0 -31 V
2346 300 M
0 31 V
0 1729 R
0 -31 V
2436 300 M
0 31 V
0 1729 R
0 -31 V
2507 300 M
0 31 V
0 1729 R
0 -31 V
2564 300 M
0 31 V
0 1729 R
0 -31 V
2613 300 M
0 31 V
0 1729 R
0 -31 V
2655 300 M
0 31 V
0 1729 R
0 -31 V
2692 300 M
0 31 V
0 1729 R
0 -31 V
2725 300 M
0 63 V
0 1697 R
0 -63 V
1.000 UL
LTb
2943 300 M
0 31 V
0 1729 R
0 -31 V
3071 300 M
0 31 V
0 1729 R
0 -31 V
3161 300 M
0 31 V
0 1729 R
0 -31 V
3232 300 M
0 31 V
0 1729 R
0 -31 V
3289 300 M
0 31 V
0 1729 R
0 -31 V
3338 300 M
0 31 V
0 1729 R
0 -31 V
3380 300 M
0 31 V
0 1729 R
0 -31 V
3417 300 M
0 31 V
0 1729 R
0 -31 V
3450 300 M
0 63 V
0 1697 R
0 -63 V
1.000 UL
LTb
1.000 UL
LTb
550 300 M
2900 0 V
0 1760 V
-2900 0 V
550 300 L
LTb
LTb
1.000 UP
1.000 UL
LT0
550 1899 M
15 0 V
16 0 V
15 0 V
15 0 V
16 0 V
15 -1 V
16 0 V
15 0 V
15 0 V
16 0 V
15 0 V
15 -1 V
16 0 V
15 0 V
15 0 V
16 -1 V
15 0 V
16 0 V
15 0 V
15 -1 V
16 0 V
15 0 V
15 -1 V
16 0 V
15 0 V
15 -1 V
16 0 V
15 -1 V
16 0 V
15 0 V
15 -1 V
16 -1 V
15 0 V
15 -1 V
16 0 V
15 -1 V
15 -1 V
16 0 V
15 -1 V
15 -1 V
16 0 V
15 -1 V
16 -1 V
15 -1 V
15 -1 V
16 -1 V
15 -1 V
15 -1 V
16 -1 V
15 -2 V
15 -1 V
16 -1 V
15 -1 V
16 -2 V
15 -1 V
15 -2 V
16 -2 V
15 -1 V
15 -2 V
16 -2 V
15 -2 V
15 -2 V
16 -2 V
15 -2 V
16 -3 V
15 -2 V
15 -3 V
16 -2 V
15 -3 V
15 -3 V
16 -3 V
15 -3 V
15 -3 V
16 -3 V
15 -4 V
16 -3 V
15 -4 V
15 -4 V
16 -4 V
15 -4 V
15 -5 V
16 -4 V
15 -5 V
15 -5 V
16 -5 V
15 -5 V
16 -5 V
15 -6 V
15 -5 V
16 -6 V
15 -6 V
15 -6 V
16 -7 V
15 -6 V
15 -7 V
16 -7 V
15 -7 V
15 -7 V
16 -8 V
15 -8 V
16 -8 V
15 -8 V
15 -8 V
16 -8 V
stroke
2148 1649 M
15 -9 V
15 -9 V
16 -9 V
15 -9 V
15 -9 V
16 -10 V
15 -9 V
16 -10 V
15 -10 V
15 -10 V
16 -10 V
15 -11 V
15 -10 V
16 -11 V
15 -11 V
15 -10 V
16 -11 V
15 -11 V
16 -12 V
15 -11 V
15 -11 V
16 -12 V
15 -11 V
15 -12 V
16 -11 V
15 -12 V
15 -12 V
16 -11 V
15 -12 V
16 -12 V
15 -12 V
15 -12 V
16 -12 V
15 -12 V
15 -12 V
16 -12 V
15 -13 V
15 -12 V
16 -12 V
15 -12 V
16 -12 V
15 -13 V
15 -12 V
16 -12 V
15 -12 V
15 -13 V
16 -12 V
15 -12 V
15 -13 V
16 -12 V
15 -12 V
16 -13 V
15 -12 V
15 -13 V
16 -12 V
15 -12 V
15 -13 V
16 -12 V
15 -12 V
15 -13 V
16 -12 V
15 -13 V
15 -12 V
16 -12 V
15 -13 V
16 -12 V
15 -13 V
15 -12 V
16 -13 V
15 -12 V
15 -12 V
16 -13 V
15 -12 V
15 -13 V
16 -12 V
15 -12 V
16 -13 V
15 -12 V
15 -13 V
16 -12 V
15 -13 V
15 -12 V
16 -12 V
15 -13 V
1.000 UL
LT1
1722 2060 M
24 -19 V
29 -24 V
29 -24 V
30 -23 V
29 -24 V
29 -23 V
29 -24 V
29 -24 V
29 -23 V
30 -24 V
29 -23 V
29 -24 V
29 -24 V
29 -23 V
30 -24 V
29 -23 V
29 -24 V
29 -24 V
29 -23 V
29 -24 V
30 -23 V
29 -24 V
29 -24 V
29 -23 V
29 -24 V
29 -24 V
30 -23 V
29 -24 V
29 -23 V
29 -24 V
29 -24 V
29 -23 V
30 -24 V
29 -23 V
29 -24 V
29 -24 V
29 -23 V
29 -24 V
30 -23 V
29 -24 V
29 -24 V
29 -23 V
29 -24 V
30 -23 V
29 -24 V
29 -24 V
29 -23 V
29 -24 V
29 -23 V
30 -24 V
29 -24 V
29 -23 V
29 -24 V
29 -23 V
29 -24 V
30 -24 V
29 -23 V
29 -24 V
29 -24 V
1.000 UL
LT2
550 1904 M
29 0 V
29 0 V
30 0 V
29 0 V
29 0 V
29 0 V
29 0 V
29 0 V
30 0 V
29 0 V
29 0 V
29 0 V
29 0 V
29 0 V
30 0 V
29 0 V
29 0 V
29 0 V
29 0 V
29 0 V
30 0 V
29 0 V
29 0 V
29 0 V
29 0 V
29 0 V
30 0 V
29 0 V
29 0 V
29 0 V
29 0 V
30 0 V
29 0 V
29 0 V
29 0 V
29 0 V
29 0 V
30 0 V
29 0 V
29 0 V
29 0 V
29 0 V
29 0 V
30 0 V
29 0 V
29 0 V
29 0 V
29 0 V
29 0 V
30 0 V
29 0 V
29 0 V
29 0 V
29 0 V
30 0 V
29 0 V
29 0 V
29 0 V
29 0 V
29 0 V
30 0 V
29 0 V
29 0 V
29 0 V
29 0 V
29 0 V
30 0 V
29 0 V
29 0 V
29 0 V
29 0 V
29 0 V
30 0 V
29 0 V
29 0 V
29 0 V
29 0 V
29 0 V
30 0 V
29 0 V
29 0 V
29 0 V
29 0 V
30 0 V
29 0 V
29 0 V
29 0 V
29 0 V
29 0 V
30 0 V
29 0 V
29 0 V
29 0 V
29 0 V
29 0 V
30 0 V
29 0 V
29 0 V
29 0 V
1.000 UL
LTb
550 300 M
2900 0 V
0 1760 V
-2900 0 V
550 300 L
1.000 UP
stroke
grestore
end
showpage
}}%
\put(2000,50){\makebox(0,0){distance $(\omega_p/c)L$}}%
\put(100,1180){%
\special{ps: gsave currentpoint currentpoint translate
270 rotate neg exch neg exch translate}%
\makebox(0,0)[b]{\shortstack{torque $|\tau_z|/(\hbar\omega_p)$}}%
\special{ps: currentpoint grestore moveto}%
}%
\put(3450,200){\makebox(0,0){ 100}}%
\put(2725,200){\makebox(0,0){ 10}}%
\put(2000,200){\makebox(0,0){ 1}}%
\put(1275,200){\makebox(0,0){ 0.1}}%
\put(550,200){\makebox(0,0){ 0.01}}%
\put(500,2060){\makebox(0,0)[r]{ 0.001}}%
\put(500,1473){\makebox(0,0)[r]{ 1e-04}}%
\put(500,887){\makebox(0,0)[r]{ 1e-05}}%
\put(500,300){\makebox(0,0)[r]{ 1e-06}}%
\end{picture}%
\endgroup
 